\documentclass[12pt]{iopart}
\usepackage{graphicx}
\usepackage[activeacute,english]{babel}
\begin{document}

\title{Bounding dissipation in stochastic models}

\author{A. Gomez-Marin$^1$, J.M.R. Parrondo$^2$ and C. Van den Broeck$^3$}

\address{$^1$ Facultat de Fisica, Universitat de Barcelona,
Diagonal 647, Barcelona, Spain \\ $^2$ Dep. F{\'i}sica
At\'omica, Molecular y Nuclear and {\em GISC}, Universidad
Complutense de Madrid, 28040 Madrid, Spain \\
$^3$ Hasselt University, B-3590 Diepenbeek, Belgium}

\ead{agomezmarin@gmail.com}

\begin{abstract}
We generalize to stochastic dynamics the exact expression  for
average dissipation along an arbitrary non-equilibrium process,
given in Phys. Rev. Lett. {\bf 98}, 080602 (2007). We then derive
lower bounds by various coarse-graining procedures and illustrate
how, when and where the information on the dissipation is captured
in models of over- and underdamped Brownian particles.
\end{abstract}




\maketitle

\section{Introduction}

Equilibrium statistical physics provides the microscopic foundation
of thermodynamics, built around the concept of {\it entropy} as the
logarithm of the phase volume. The theory has been extended to the
regime of linear irreversible thermodynamics by identifying the {\it
entropy production} in the regime of linear response \cite{onsager,onsager2,prigogine,prigogine2}. There exists to date no general theory
covering the far from equilibrium situations. However, recent
results known as fluctuation
\cite{fluctuation,fluctuation2,fluctuation3,fluctuation4,fluctuation5}
or work \cite{work,worka,workb,workc,work2,work2a,work3,work4,work5} theorems point to the
existence of exact equalities valid independent of the distance from
equilibrium. These equalities involve fluctuations in work or
entropy production. For the average of these quantities, they reduce
to inequalities, in agreement with the second law of thermodynamics.
For example, the Jarzynski equality states that $\langle \exp(-\beta
W) \rangle = \exp (-\beta \Delta F)$, where $W$ is the work needed
to bring a system, in contact with a heat bath at temperature $T$
($\beta^{-1}=k_B T$), from one initial state prepared in equilibrium
to another one and $ \Delta F$ is the difference in free energy  of
these states (see  \cite{work2} for a more precise discussion). By
the application of Jensen's inequality, one finds $\langle W \rangle
\ge \Delta F$.

While the work and fluctuation theorems are certainly intriguing
results of specific interest for the study of small systems, they
provide no extra information on the average value of work and
entropy production. Recently however, the microscopically exact
value of the average work has been obtained in a set-up similar to
that of the work theorem \cite{diss}.  The system is described by a
Hamiltonian $H(\Gamma,\lambda)$, where $\Gamma=(\{q\},\{p\})$ is a
point in phase space, representing all position and momentum
variables, and $\lambda$ is an external control parameter (for
example the volume or an external field). The system is perturbed
away from its initial canonical equilibrium  by changing the control
parameter according to a specific schedule, from an initial to a
final value. This involves a certain amount of work $W$, which is a
random variable due to the randomness of the initial state. By
repeating the experiment (or by solving Liouville's equation) one
can, in principle, evaluate the probability density $\rho(\Gamma;t)$
for the system to be in a specific micro-state  $\Gamma$   at a
specific (but otherwise arbitrary) intermediate time $t$ during the
transition. Furthermore one considers the time-reversed scenario, in
which the system starts in canonical equilibrium at the final value
of the control parameter, and the latter is changed following the
time-reversed schedule. We will use the superscript ``tilde'' to
refer to such time-reversed corresponding quantities. Then one
measures the phase space density ${\tilde{\rho}}(\tilde{\Gamma};t)$,
at the moment when the control parameter reaches the same value as
the one considered in the forward experiment  (so $t$ here stands
for the forward time and $\tilde{\Gamma}=(\{q\},\{-p\}$). The
dissipated work $\langle W \rangle -\Delta F$, which is the
``unknown positive quantity'' appearing in the second law, is then
found to be given by the following explicit result:
\begin{eqnarray} \label{maineq00}
\langle W \rangle -\Delta F = k_BT\int d\Gamma \;\rho(\Gamma;t) \ln
\frac{\rho(\Gamma;t)}{\tilde{\rho}(\tilde\Gamma;t)}=k_B T D(\rho
||{\tilde{\rho}}).
\end{eqnarray}

$D(\rho ||{\tilde{\rho}})$ is the relative entropy, also called
Kullback-Leibler distance \cite{cover}. It is a positive  quantity,
in agreement with the second law. While the above result is exact
and fully  reveals the microscopic nature  of the  dissipation, it
may appear to be of little practical interest. Indeed, it requires
{\it full statistical information} on {\it all} the microscopic
degrees of freedom of the system (even though only at one particular
time).  This stringent requirement is obviously on par with the
generality of the above result, which is valid however far the
system is perturbed away from equilibrium. The perturbation could
therefore imprint its effect on all the degrees of freedom and their
full statistical information would be required to reproduce the
corresponding dissipation.

One main purpose of this paper is to tune the above result to
situations in which only a limited number of degrees of freedom are
either relevant or available. An important class of such systems,
notable for its accurate description of mesoscopic phenomena in
physics, chemistry and biology, are stochastic models such as the
Master Equation or the Langevin equation \cite{vankampen}. 
The application of Eq.~(\ref{maineq00}) to systems described
by stochastic dynamics is not obvious. Instead,
we will derive in the next section a general and
exact result applicable to stochastic systems by
rewriting Eq. (1) in an alternative form, as an
integral over paths. A simple argument to derive
this formulation for Hamiltonian dynamics goes as
follows.

Since the microscopic dynamics is completely deterministic, the
specification of an elementary phase space volume $d\Gamma$ around
the position $\Gamma$ at time $t$ is equivalent to the
identification of an elementary ensemble of paths
$\cal{D}({\mbox{path}})$ surrounding the phase space trajectory
going through $\Gamma$ at time $t$. The probability to select a path
inside this bundle (of constant cross
section, $d\Gamma$ being preserved following
Liouville's theorem)
will be denoted by $\cal{D}({\mbox{path}})
{\cal{P}(\mbox{path})}$, where $ {\cal{P}(\mbox{path})}$ is the
probability density in function space. Similarly, one defines the
density $\widetilde{\cal{P}}(\widetilde{\mbox{path}})$ for the
time-reversed schedule.  In Sec. 2 we will prove that
Eq.~(\ref{maineq00}) can be rewritten as follows:
\begin{eqnarray}
\langle W \rangle -\Delta F &=& k_BT\int \cal{D}({\mbox{path}}) {\cal
P}(\mbox{path})\ln \frac{ {\cal P}(\mbox{path})}{\widetilde{{\cal
P}} (\widetilde{\mbox{path}})}\nonumber \\ &=& k_BTD({{\cal
P}(\mbox{path})}||{\widetilde{{\cal P}}
(\widetilde{\mbox{path}})})\label{maineq}.
\end{eqnarray}

The above expression is in principle valid only if ``path'' stands
for the microscopic (and hence deterministic) trajectory of the
system, including information about every degree of freedom. If only
partial information about this trajectory is taken into account, the
relative entropy  is typically reduced and one cannot derive an
equality for the dissipation, but just a {\it lower bound}. However,
we will show in the next section that in the path  description, not
all variables are always needed. This will be a welcome
simplification since a detailed description of the microstate of the
system is rarely available, especially if the system contains many
"thermal degrees" of freedom. Furthermore, we will also focus on the
case of missing information, both at the level of the path and at
the level of variables. In this case one can produce lower bounds
for the dissipation. This issue has been briefly discussed in the
incipient letter \cite{diss}, but will be addressed here in greater
detail and broader generality.

The layout of this paper is as follows. We first
prove in Sec. \ref{sec:dis}, that Eq. (\ref{maineq}) is also valid,
{\em as an equality}, for stochastic dynamics and discuss the
relation between this result and Crooks' theorem. We next
investigate in a number of experimentally relevant examples, how relative entropy, and
consequently the estimated dissipation, decreases when only partial
or coarse grained information on the system is taken into account.
Such coarse-graining can be applied to the measurement in time, or
to the choice of variables. We will present illustrations for both
cases. First we calculate in Sec.~\ref{sec:over} the lower bound for
dissipation upon coarse graining in time for an overdamped Brownian
particle in a time-dependent moving harmonic potential  \cite{ritort,gaspard2007}. We discuss the
convergence to the exact dissipation as the number of measurement
points of the stochastic trajectory increases. Next we turn in Sec.~\ref{sec:under} to an
underdamped Brownian particle in a harmonic potential.
We illustrate how the information about the dissipated work involved in quenching the
potential oscillates, in a single time measurement, between position and
momentum variables and eventually is irreversibly lost into the heat bath
variables. Finally, we consider in Sec. \ref{sec:under2} a Brownian
particle in a quenched harmonic potential in contact to a heat bath
via a second Brownian particle. The information about the
dissipation is then found to channel back and forth between the 4 degrees
of freedom, position and momentum of both particles, in a very
intricate and  intriguing way.

\section{Dissipation in stochastic dynamics}\label{sec:dis}

The derivation (for stochastic models), interpretation and application of the main result Eq.~(\ref{maineq}) heavily
relies on  two basic properties of the relative entropy, namely Stein's lemma
and the chain rule  \cite{cover}, which we now review.

Stein's lemma gives a more precise operational meaning to the
relative entropy. The
relative entropy $D(p||q)$ between two different distributions
$p(x)$ and $q(x)$ quantizes the likeliness for independent samplings
from $p(x)$ to be statistically identified as samplings from $q(x)$.
More precisely, Stein's lemma states that the chance for mistakenly
attributing a series of $n$  samplings from $p(x)$ to $q(x)$ decreases (at best) exponentially as  $e^{-n D(p||q)}$. Hence, the
identification of the statistical source becomes exponentially more
difficult when the relative entropy decreases. Note that the
relative entropy is not a symmetric function of its arguments. This
is consistent with the fact that  the difficulty to distinguish $p$
from $q$ depends on whether the samplings come from $p$ or $q$.

As applied to our basic result, Eq. (\ref{maineq00}) or
(\ref{maineq}), we conclude that the dissipated work is essentially
related to the difficulty for {\it distinguishing the arrow of time}: the
dissipated work will be small (or large) in the forward experiment,
when realizations of that process can be easily (or hardly) confused
with those appearing in the backward process. Typically, the
dissipated work is extensively large for macroscopic systems when
operating away from the quasi-static regime, and the arrow of time
is clearly apparent. Close to the quasi-static regime with small
dissipation, the system is near equilibrium at each instant of time
and both snapshots and runs in the forward or backward experiments will look very much
alike. When operating away from the quasi-static regime in
sufficiently small systems, it may still take several runs to
clearly statistically distinguish forward from backward runs. The
dissipated work is clearly positive but may be small or comparable
to $k_BT$.

Next we turn to the chain rule. Consider two random variables $X$ and $Y$.
The relative entropy between two different distributions $p(x,y)$
and $q(x,y)$ can be written as
\begin{eqnarray}
\fl  D\left( p(x,y)||q(x,y)\right) & = & \int
dx\,dy\,p(x,y)\ln\frac{p(x,y)}{q(x,y)}\nonumber \\  &=& \int
dx\,dy\,p(x,y)\ln\frac{p(y|x)p(x)}{q(y|x)q(x)}  \nonumber \\
& = & D\left(p(x)||q(x)\right) + \int\! dx \,p(x)\!\int\! dy\,
p(y|x)\ln\frac{p(y|x)}{q(y|x)},
\end{eqnarray}
a result referred to as the chain rule for the relative entropy. Since the
last term in the r.h.s is  non-negative, one concludes
\begin{equation}
D\left( p(x,y)||q(x,y)\right)\ge D\left(p(x)||q(x)\right).
\end{equation}
This inequality has a simple intuitive explanation. The relative
entropy is a measure of the distinguishability of the two
probability distributions. It is obvious that the distinction will
be easier to make when considering the statistical information of
the two variables rather than only one of them. Next, we mention the
following special cases of the chain rule. If $X$ and $Y$ are
independent, we have
\begin{equation}
D\left( p(x,y)||q(x,y)\right)=D\left( p(x)||q(x)\right)+D\left(
p(y)||q(y)\right),
\end{equation}
However, note that, in general, the sum of $D\left( p(x)||q(x)\right)$ and $D\left(
p(y)||q(y)\right)$ can be either bigger or smaller than $D\left( p(x,y)||q(x,y)\right)$.

If $X=f(Y)$, $f$ being a one-to-one function, then one of the
variables does not add any information about the other one, hence
\begin{equation}\label{rei}
D\left( p(x,y)||q(x,y)\right)=D\left( p(x)||q(x)\right)=D\left(
p(y)||q(y)\right).
\end{equation}

This last observation  ---that the addition of variables which are
functions of the existing ones, leaves the relative entropy
invariant--- provides a rigorous derivation of Eq.~(\ref{maineq})
from Eq.~(\ref{maineq00}) for Hamiltonian dynamics. Indeed, since
Hamiltonian dynamics is purely deterministic, one can specify,
without changing the value of the relative entropy, the micro-state
$\Gamma_i$ of the system at as many additional measurement points in time  $t_i$, $ i=1,...,n$, as one likes in Eq.~(\ref{maineq00}):
\begin{eqnarray} \label{maineq001}
\langle W \rangle -\Delta F = k_BT\int \prod_{i=1}^{n}
d\Gamma_i
 \;\rho(\{\Gamma_i;t_i\}) \ln
\frac{\rho(\{\Gamma_i;t_i\})}{\tilde{\rho}(\{\tilde{\Gamma}_i;t_i\}}.
\end{eqnarray}
In the continuum limit covering the entire time interval (with $n
\rightarrow \infty$), one thus converges to the path integral
formulation given in Eq.~(\ref{maineq}). This expression, while
containing redundant information from the point of view of
Hamiltonian dynamics, has the important advantage (see below) that
it is also exact and formally identical (i.e., the path is now in
terms of the reduced set of stochastic variables) in its application
to stochastic systems. Furthermore, in the latter case, the path
formulation is no longer redundant since the trajectory captures
information about the eliminated degrees of freedom.

The straightforward  application of the chain rule in Eq.
(\ref{maineq00}) or (\ref{maineq}) now leads to the following result:
\begin{equation} \label{maineq01}
\langle W \rangle -\Delta F \geq k_BT D({\cal{P} }(x)||\tilde{\cal{P}}(\tilde {x})),
\label{bound}
\end{equation}
where $x$ is any partial information on the path followed by the
system with corresponding probability ${\cal{P}}(x)$. The  variables
$x$ can reflect a reduction in the number of variables, a measurement of these variables in a coarse grained fashion, or
the reduction to  partial or even punctual information in time along
the path.
The above formula is quite useful, since it improves on the second
law statement $\langle W \rangle -\Delta F \geq 0$, with whatever
information is available. Furthermore, the formula in principle
allows one to identify which are the relevant variables or degrees
of freedom whose time symmetry breaking is the most relevant to
estimate the dissipation. The following sections  will be devoted to
illustrate these issues in detail on several explicit examples.

The chain rule also reveals a most interesting relation  between
the probabilities for paths and probabilities for work by comparing
the general result for the average dissipation Eq. (\ref{maineq00})
or (\ref{maineq}) with the Crooks theorem. The latter has been
proved for both stochastic \cite{crooks} and Hamiltonian systems
\cite{work5} and states that:
\begin{equation} k_BT\ln \frac{P(W)}{\widetilde P(-W)}=W-\Delta
F\label{crooks0},
\end{equation}
where $P(W)$ and $\widetilde P(W)$ are the probability distributions
for the work in the forward and the backward process, respectively.
From this  theorem, one immediately obtains the following
expression for the average dissipation:
\begin{equation} \langle
W\rangle -\Delta F=k_BT D(P(W)||\widetilde P(-W)).
\label{crooks}\end{equation}
By comparison with
Eq.~(\ref{maineq}), we conclude that
\begin{equation} \label{dwdp}
D( {\cal P} (\mbox{path}) || \widetilde{{ \cal
P}}(\widetilde{\mbox{path}})) = D(P(W)||\widetilde P(-W)).
\end{equation}
Crooks theorem thus implies that the time asymmetry in the
probability distribution of the work fully determines the average
dissipation. This is a surprising relationship with important
consequences.  From the chain rule for the relative entropy one
would expect that the relative entropy for the paths, which contains
all the information on the process under consideration, would be
bigger than that contained in a single variable, namely the work.
However, the two relative entropies are equal, indicating that the
statistical information about the work in the forward and backward
processes accounts for {\em every} appearance of the arrow of time.

The foregoing intriguing conclusion can be turned around   to
provide a general derivation of Eq.~(\ref{maineq}), as applied to
stochastic processes. Indeed, the work obviously depends only on the
dynamic variables that are interacting with the external device
during the process. Hence it is enough to know the (statistical)
behavior of these variables to reproduce the statistics of the work,
and hence the average dissipation. In principle, this even needs not
be all the variables, but only those that matter for the energy
exchange. In that case, trajectory information of these and only
these variables, along the whole (both forward and the backward)
process, is enough to account for the total average dissipation. In
particular, if a stochastic model provides the exact description of
a system in its interaction with an external device, one needs only
the path information of these variables. Eq.~(\ref{maineq}) is thus
valid for the stochastic model with the path determined in terms of
the corresponding stochastic variables. As a corollary, we note that
bath variables which are replaced (in some ideal limit) by a
stochastic perturbation, will not appear in the ``path", which is
in terms of the trajectory of the stochastic system only.

The above argument forms an alternative derivation for stochastic
models, complementing a more standard proof which runs as follows.
We distinguish, in the path integral of Eq.~(\ref{maineq}) involving
all microscopic variables, the integration over fast microscopic
variables and the relevant mesoscopic stochastic variables. The idea
is that the microscopic variables can be integrated out, leaving a
path integral over the stochastic variables only. This will be the
case if the dependence on the micro-variables disappears under
logarithm in the ratio of the probability of a path and its
time-reverse. They can then be further integrated out, leaving only
stochastic variables. The usual scenario implies a limit involving a
separation of time scales: the micro-variables assume
instantaneously an equilibrium distribution (which is the same in
both forward and backward process and hence drops out) for the given
values of the slow stochastic variables. Being all the time at the
instantaneous equilibrium implies that they do not carry any
time-asymmetry and thus, not surprisingly, do not contribute to the
dissipation. Notice however that our first derivation, based on
Eq.~(\ref{dwdp}), indicates that the separation of the two time
scales may not be necessary, since only the variables that determine
the work will be required.  In particular this derivation puts no
limitation on the nature of the stochastic process, other than that
it be consistent with (derivable from) the microscopic dynamics. In
particular, the process does not have to be Gaussian nor Markovian.

We need to make some additional remarks on the interpretation of
Eq.~(\ref{maineq}) for stochastic processes.  First, we have to
recall that, in the derivation of the above result, it is assumed
that the system starts in canonical equilibrium in both forward and
backward scenario.  Hence the integral over the paths, whether
microscopic or stochastic, has to be performed over initial
canonical distributions. Second, we note that in the switch to a
stochastic process, we need to consider the distribution of paths
during the entire duration of the experiment. Only for Hamiltonian
dynamics does the phase space density and one particular instant of
time carry all the information on the dissipation during the entire
experiment. Notice also that the general bound (\ref{bound}), for an
arbitrary description of the system given by the set of variables
$x$, does not follow from Crooks theorem. We do need the microscopic
results Eqs.~(\ref{maineq00}) or (\ref{maineq}) to conclude that any
``additional'' information contained in $x$ does not lead to an
increase of the relative entropy. In other words, an overestimation
of the dissipation  via the relative entropy is excluded.

We finally put Eq.~(\ref{maineq}) in the context of various results
from the literature. The importance of relative entropy in
nonequilibrium statistical mechanics has been the object of general
discussions, both in the context of classical Hamiltonian mechanics
\cite{Mackey} and quantum mechanics \cite{Vedral}.  
A result for the average work is also derived in \cite{jarzynski2006}, but the focus of such paper is on the interpretation of
the Jarzynski equality, and the connection with the relative entropy
(and with its extremely useful properties) is not made. The above
expression (\ref{maineq})  for the mean dissipation is consistent
with earlier results for Markovian dynamics \cite{luo}.
More recently,  arguments have been produced to show that
\begin{equation}
k_B   \ln \frac{\cal{P}(\mbox{path})}{\widetilde{\cal{P}} (\widetilde{\mbox{path}})}
\end{equation}
is the correct expression for the {\it path dependent} entropy
production in Markovian stochastic systems
\cite{sekimoto,kurchan,qian,maes,wu}, see especially the early works by Maes
\cite{fluctuation5} and by Crooks \cite{crooks,crooksthesis}. The
connection with Eq.~(\ref{maineq}) is made by observing that the
dissipated work is evacuated to the heat bath as heat so that
Eq.~(\ref{maineq}) is equal to average entropy production divided by
the temperature. Note that the average entropy production is always
positive, while the path dependent expression can have any sign.
More recent discussions include  general arguments based on coarse
graining \cite{maes}, Langevin equations \cite{seifert2005},
stationary stochastic processes \cite{porporato}, and an
experimental verification for dragged Brownian particles
\cite{gaspard2007}. A similar formula has also been proposed for
dynamical systems \cite{gaspard2004}, to characterize the time
asymmetry of the Sinai-Kolmogorov entropy.

\section{Overdamped Brownian particle: coarse graining in time}\label{sec:over}

Referring to Eq. (\ref{maineq}) and our  preceding discussion
concerning the variables that need to be included in the path
integral, it would be welcome to have a simple explicit example in
which all the calculations can be done analytically.  In this
section we present such a case which is moreover of experimental
relevance, namely the case of an overdamped Brownian particle
subject to  a moving time-dependent harmonic potential:
\begin{equation}\label{mhp}
V(x,t)=\frac{k}{2}(x-ut)^2.
\end{equation}
This same example will also provide a simple  illustration of the
chain rule as applied to coarse graining in time.

\subsection{Stochastic energetics for a Langevin equation with time dependent potential}

 The time evolution of the position variable  $x$ of  the overdamped particle obeys the following Langevin equation
\begin{equation}\label{lang}
\dot{x}=-\partial_x V(x,t)+\xi(t).
\end{equation}
$\xi(t)$ is a Gaussian white noise, with $\langle
\xi(t)\xi(t')\rangle=2T \delta(t-t')$. For simplicity of notation,
we have absorbed the friction coefficient in the time unit and the
Boltzmann constant $k_B$ in the definition of temperature. Before
proceeding to the relation between dissipation and relative entropy,
we review the salient features of the energy balance.

Our starting
point is conservation of total energy during an experiment from
initial time $t_0$ to final time $t_f$. Since the particle is
instantaneously thermalized at the constant temperature $T$ of the
heat bath, its change in energy is equal to its change in potential
energy $\Delta V=V(x_f,t_f)-V(x_0,t_0)$. The latter must be
equal to the amount of work $W$ exerted by the external force
(sometimes called the injected work) minus the heat $Q$ delivered to the
heat bath (also referred to as dissipated heat to the environment)
\begin{equation} \label{1stLaw}
\Delta V=\int_{t_0}^{t_f} \frac{dV}{dt}dt=
\int_{t_0}^{t_f} \frac{\partial V}{\partial t}dt
+\int_{t_0}^{t_f} \frac{\partial V}{\partial x} \dot{x}dt=W-Q.
\end{equation}
From such energy balance (or first law at the level of stochastic quantities), the fluctuating heat and work can be identified \cite{sekimoto};
the rate of heat dissipated to the heat bath is given by
$\dot{Q}=-\partial_x {V} \dot{x}$, while the work done per unit time in moving the external potential is $\dot{W}=\partial_t{V}$.
These quantities depend on the actual realization of
the stochastic trajectory $x(t)$. Thus heat and work are random variables.
The fact that injected work and dissipated heat differ by the energy stored in the particle
has important consequences for their large deviation properties for
asymptotically large times when the latter energy is unbounded. The
fluctuation theorem has therefore to be carefully reconsidered
~\cite{farago,vanzon,blickle,hfbt,visco,maes2006,joubaud}.

We will be concerned here with the average work, in which case large
deviation issues are irrelevant.
Using the explicit expression of the potential (\ref{mhp}), one finds
\begin{eqnarray}
\langle W \rangle &=& \left\langle \int_{t_0}^{t_f}\frac{\partial
V(x,t)}{\partial t}  dt  \right\rangle = \left\langle
\int_{t_0}^{t_f} dt k(x-ut)(-u)  \right\rangle \nonumber \\ &=& u
\int_{t_0}^{t_f} dt \langle \dot{x}(t)-\xi(t) \rangle =u [\langle
x(t_f)\rangle -\langle x(t_0) \rangle].
\end{eqnarray}

On the other hand, the average of equation (\ref{lang}) yields the
following exact closed equation for the average position
\begin{equation}\label{langeq}
\dot{\langle x \rangle} = -\langle \partial_x V \rangle=-k(\langle x \rangle -ut).
\end{equation}
The solution reads:
\begin{equation}\label{mean}
{\langle x(t_f) \rangle} =e^{-k(t_f-t_0)}\langle x(t_0) \rangle+\frac{u}{k} [kt_f -1-e^{-k(t_f-t_0)}(kt_0-1)].
\end{equation}
We take the initial time $t_0\equiv 0$. Then, since the system must be prepared initially in equilibrium, from the evolution equation it is obvious that $\langle x(t_0) \rangle=0$.

The translation of the harmonic
potential minimum does not change the free energy of the system,
$\Delta F=0$.  We then obtain that the dissipated work, being exactly
equal to average work, is given by
\begin{equation} \label{Wdiss}
\langle W_{diss} \rangle \equiv \langle W \rangle- \Delta F= \frac{u^2}{k} (kt_f +e^{-kt_f}-1).
\end{equation}
In the sequel, we will illustrate how this result is approached from below as
we obtain more the information on the paths by an increasing number
of measurements in time.

\subsection{Coarse-graining in time}

It is obviously impossible to numerically or experimentally measure
with infinite precision the full trajectory of a particle.  Instead,
its position $x(t)$ can be recorded at a finite number of
measurement points in time. This information loss about the path can
be viewed as a coarse-graining (in time). By replacing the path
integral by the corresponding finite sum, one finds an approximate
value for the dissipation. However, as mentioned before, one gets
more: this result, and in fact any result obtained through
coarse-graining, constitutes a rigorous lower bound.
The calculation which we are about to perform will tell us how fast
this bound converges to the exact value.

\begin{figure}
\begin{center}
\includegraphics[angle=0, width=12cm]{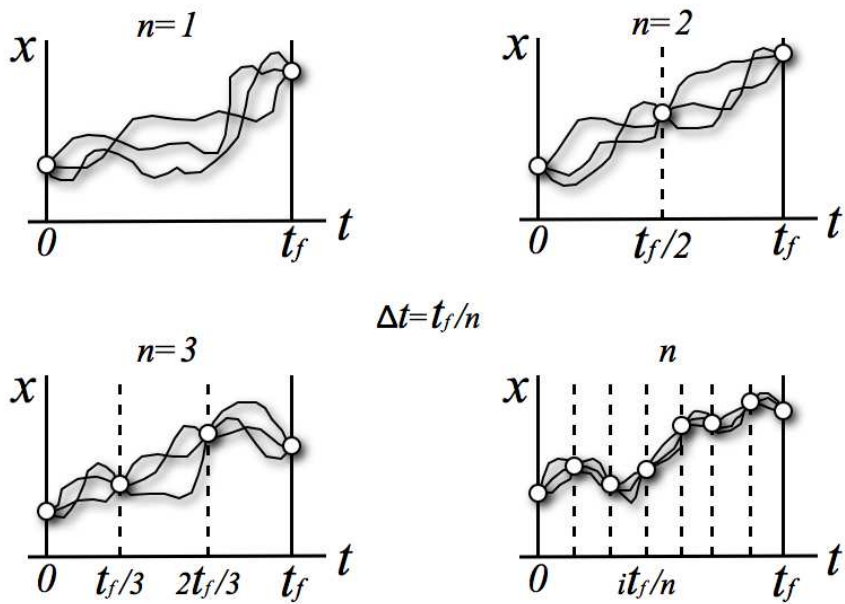}
\caption{Sketch of the $n$-slicing procedure in which the full
trajectory of the particle is not measured but only its position
after time intervals of duration $\Delta t =t_f/n$, where $t_f$ is the total
time of the experiment. } \label{scheme}
\end{center}
\end{figure}

For simplicity we will consider that the coarse graining is into $n$
equal divisions $\Delta t\equiv t_f/n$ of the total time duration $t_f$. Therefore, in this $n$-slicing procedure, the full trajectory of the particle  is not measured but only its position after time intervals of duration $\Delta t$. See figure (\ref{scheme}).
The probability for a discretized path
can be easily evaluated since the process is Markovian and Gaussian.
Let us denote by $p(x_{i+1}|x_i)$ the conditional probability for
jumping from a point $x_i$ at time $t_i$ to a point $x_{i+1}$ at
time $t_i+\Delta t$, and let $p^{eq}_{0}$ be the initial equilibrium
distribution.  The probability $\mathcal{P}$ of the $n$-sliced discretized path
$\vec{x}\equiv[x_0,\;x_{1},\;...,\;x_{i},\;...,\;x_{n-1},\;x_{f}]$
 is then given
by
\begin{equation}\label{markovian}
\mathcal{P}(\vec{x})\equiv\mathcal{P}([x_0,\;x_{1},\;...,\;x_{i},\;...,\;x_{n-1},\;x_{f}])=
p^{eq}_{0}(x_0) \prod_{i=0}^{n-1} p(x_{i+1}|x_i).
\end{equation}
An analogous expression is valid for the backward path and
probability, with superscript ``tilde'' again referring to time
reversed excursion. The central quantity to evaluate is the
following coarse-grained relative entropy  $I_n$ (multiplied by $T$,
since we want to compare with the dissipated work, and having
absorbed $k_B$ in its units):
\begin{equation} \label{sumIn}
I_n \equiv T D(\mathcal{P}(\vec{x})||\tilde{\mathcal{P}}(\tilde{\vec{x}}))
=
T\left\langle  \ln \frac{p^{eq}_0(x_0)}{p^{eq}_f(x_f)} \right\rangle +
T\sum^{n-1}_{i=0} \left\langle \ln \frac{p(x_{i+1}|x_i)}{\tilde{p}(x_i|x_{i+1})} \right\rangle.
\end{equation}
The brackets $\langle ... \rangle$ refer to the average performed with the forward distribution $\mathcal{P}$.

The next step is to find the general expression for $p(x_{i+1}|x_i)$
and $\tilde{p}(x_i|x_{i+1})$. Since the Langevin equation that
describes the    dynamics is linear, the conditional probabilities
are Gaussian distributions
\begin{equation} \label{pF}
p(x_{i+1}| x_{i})= \frac{1}{\sqrt{2\pi \sigma^2}} \exp \left[-\frac{
(x_{i+1}- \langle x_{i+1}\rangle_{x_i} )^2
  }{2\sigma^2}\right]
\end{equation}
and
\begin{equation} \label{pB}
\tilde{p}(x_{i} |x_{i+1})= \frac{1}{\sqrt{2\pi \sigma^2}}  \exp \left[-\frac{
(x_{i} -\langle \tilde{x}_{i}\rangle_{x_{i+1}})^2
  }{2\sigma^2}\right].
\end{equation}

From equation (\ref{mean}) (applied for final and initial times
$t_{i+1}$ and $t_i$, respectively, and with the appropriate initial
condition) the conditional averages are found to be
 \begin{equation}
\langle x_{i+1}\rangle_{x_i} =x_i e^{-k\Delta t}+\omega+ \eta \;t _i
\end{equation}
and
 \begin{equation}
\langle \tilde{x}_{i}\rangle_{x_{i+1}} =x_{i+1} e^{-k\Delta t}-\omega+\eta \;t _{i+1}
\end{equation}
where
\begin{equation}
\omega \equiv \frac{u}{k}(e^{-k \Delta t}+k\Delta t -1), \; \; \eta \equiv u(1-e^{-k \Delta t}).
\end{equation}
Similarly, one can multiply the Langevin equation by the position
$x$ and then take averages. This leads to the following equation for
the variance $\sigma^2 \equiv \langle x^2\rangle-\langle x \rangle^2$:
\begin{equation}
\frac{1}{2}\frac{d}{dt}\sigma^2=-k\sigma^2 +T,
\end{equation}
which yields (conditional variances starting at zero value)
\begin{equation}
\sigma^2=\frac{T}{k}(1-e^{-2 k \Delta t}),
\end{equation}
for both (forward and backward) cases.

In order to obtain $I_n$, we insert the above conditional
probability distributions in Eq.~(\ref{sumIn}), work out the squares
and arrange the averages. The final result can most revealingly be
written in terms of the duration of the experiment $t_f$ and the
final position $z_0\equiv u t_f$. After some algebra, one finally
gets
\begin{equation}\fl \label{In!}
I_n=\frac{z^2_0}{k t^2_f}  \frac{e^{-kt_f} -(2n+1) + (1-kt_f)e^{-kt_f\frac{n-1}{n}}
+(2n-1+kt_f e^{-kt_f})e^{ \frac{kt_f}{n}}
   }{1+e^{\frac{kt_f}{n}}}.
\end{equation}
We also mention explicitly the results for $n=1$ and $n=2$:
\begin{equation}
I_1=\frac{z^2_0}{k t^2_f}  \frac{e^{-kt_f} +e^{kt_f}-2
   }{1+e^{kt_f}}
\end{equation}
and
\begin{equation}
I_2=\frac{z^2_0}{k t^2_f}  \frac{e^{-kt_f} +e^{-\frac{kt_f}{2}}+3e^{\frac{kt_f}{2}}-5
   }{1+e^{\frac{kt_f}{2}}}.
\end{equation}

\begin{figure} [t]
\begin{center}
  \includegraphics[angle=270, width=7.7cm]{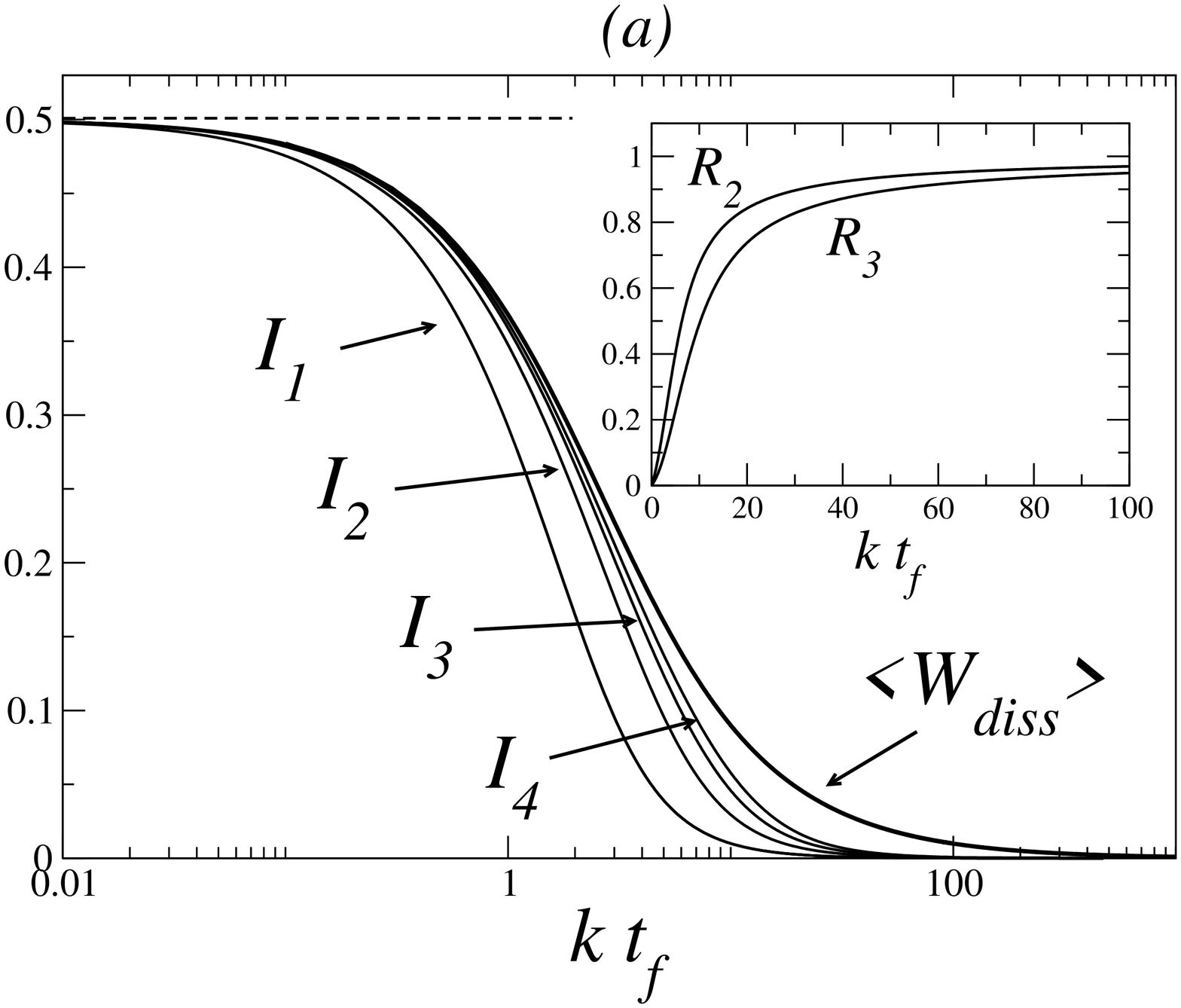}
   \includegraphics[angle=270, width=7.7cm]{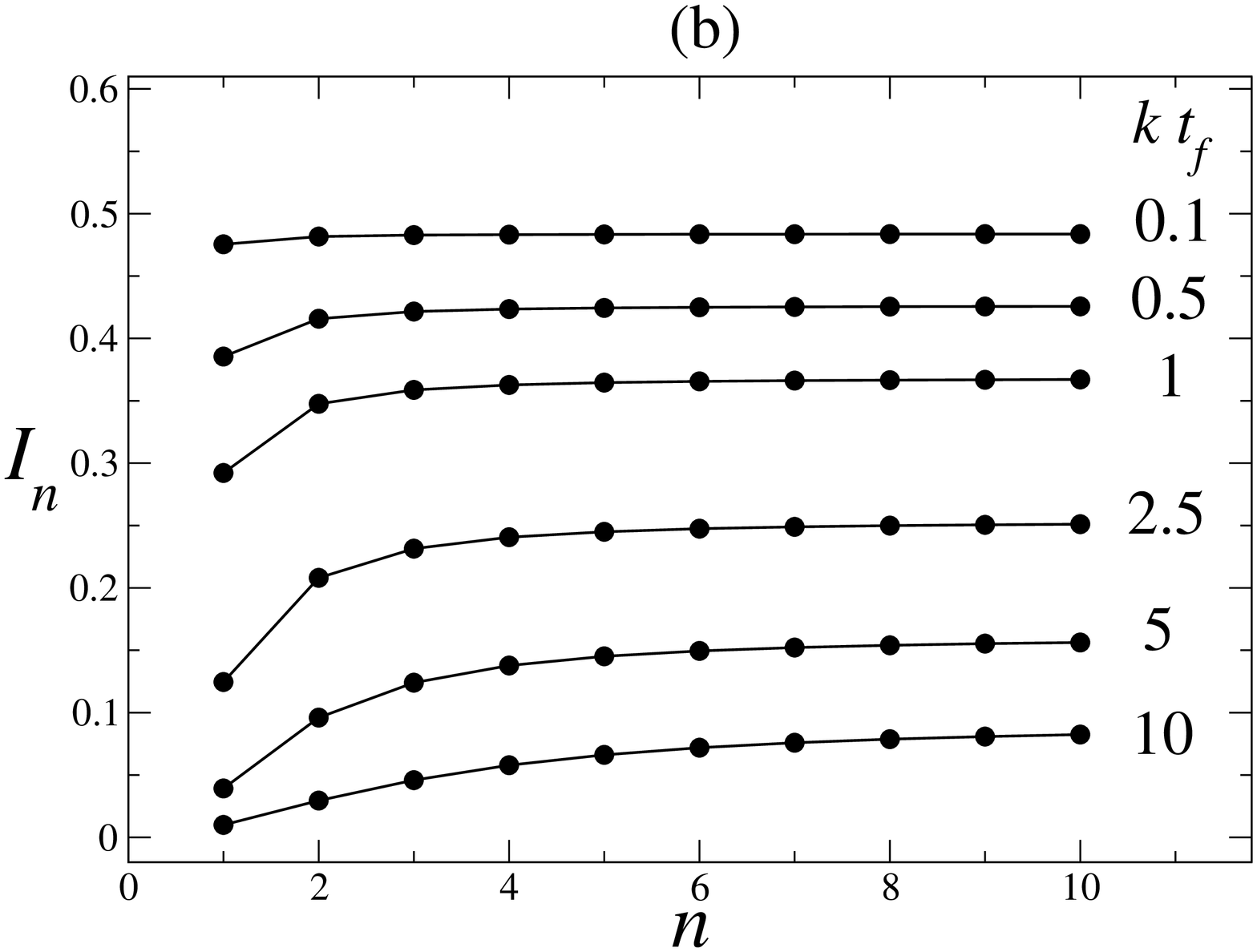}
  \caption{(a) Plot of $I_n$ (for $n=1,2,3,4$ and $I_\infty=\langle W_{diss}\rangle$), as a function of the ratio of characteristic times $k t_f $. We have scaled out
  the prefactor $k z^2_0$. Note that  $I_n$ is always a lower bound to
  $\langle W_{diss}\rangle$ and converges to the  irreversible instantaneous
  quench value (dashed line) and to the quasi-static limit (zero value) for  $k t_f \rightarrow 0$ and
  $k t_f \rightarrow \infty$, respectively. Inset:  the relative error
  $R_n=(\langle W_{diss}\rangle-I_n)/\langle W_{diss}\rangle$ increases as a function of
  $k t_f $.
  (b) Plot of $I_n$ (for different values of $kt_f$) as a function of the number of time divisions $n$ of the trajectory.}
\label{boundsIn}
\end{center}
\end{figure}

First note that in the limit $n\to \infty$ one finds (cf.
Eq.~({\ref{Wdiss}))
\begin{equation} \label{I8}
I_{\infty}= \frac{ z^2_0}{kt_f^2}(kt_f +e^{-kt_f}-1)=\langle W_{diss}\rangle.
\end{equation}
Hence the exact dissipation is, as anticipated, recovered in the
limit of the continuous path description. Using the same procedure,
one can show that this result  remains valid  for a general time
dependent potential (see the Appendix).

We now turn to the main question of interest here. How is the
convergence of  $I_n$ to $\langle W_{diss}\rangle$? First,   one can
verify that, for any value of the system's parameters, $I_n$ is
always a lower bound for the total dissipation:
\begin{equation}
\langle W \rangle \geq  I_n \geq 0.
\end{equation}
Next, as is apparent from the explicit result, the convergence of
$I_n$ to $\langle W_{diss}\rangle$ depends only on the ratio of the
time of the experiment $t_f$ over the relaxation time $1/k$ in the
harmonic potential. In figure (\ref{boundsIn}.a) we plot $I_1$ up to
$I_4$, as a function of $k t_f$. The convergence is surprisingly
good. For example, for $k t_f=1$, the error in $I_2$ (single
intermediate measurement point, plus the initial and the final
points, which are always measured) is only a few percent. We also
study in figure (\ref{boundsIn}.b) the evolution of $I_n$, for
different values of $kt_f$, as the number of measured points
increases. Note that the biggest jumps in $I_n$ occur from $n=1$ to
$n=2$, after which the bound quickly saturates and slowly approaches
the total mean dissipated work.

In the  limit $u \rightarrow 0$ or $k t_f \rightarrow
\infty$ (very slow translation of the potential), one recovers the  quasi-static result of zero dissipated work. Note however that the relative
rate of convergence becomes quite bad in this limit (cf. inset in
figure (\ref{boundsIn}.a)). On the other hand, the fit is perfect in
the limit of the irreversible quench, in which the potential is
instantaneously switched to its new position. This corresponds to
the limit $u \rightarrow \infty$ or $k t_f \rightarrow 0$. One
finds
\begin{equation}
I _n (t_f \to 0)=\langle W_{diss} \rangle (t_f \to 0)  =\frac{1}{2}kz^2_0\;\;\;\;\;\; \forall n.
\end{equation}
The dissipated work is exactly equal  to the average work done in
instantaneously placing the particle in the shifted potential.

\section{Underdamped Brownian particle: coarse graining in the space of
variables}\label{sec:under}

According to Eq.~(\ref{maineq00}),  the average dissipated work is
obtained from a single time measurement of forward and backward
statistics of  the full system. Eq.~(\ref{maineq}) provides a
complimentary result, since the measurement of some (e.g. heat bath)
variables can be avoided and the average dissipated work is still
obtained if the reduced set of stochastic variables are measured along the
whole time track of the experiment. In the previous section, we
discussed the effect of  coarse graining in time for the measurement
of  the single relevant variable at hand, namely the position of the
overdamped Brownian particle. In this section, we address the
additional question about the role of specific variables (or degrees of freedom) in
revealing the dissipation.

We naturally turn for the illustration of this point  to underdamped
Brownian particles in a harmonic potential since both position and
momentum of the particle are relevant. Instead of considering a
moving harmonic potential with fixed strength, we turn to another
experimentally significant scenario of a non-moving harmonic
potential undergoing an instantaneous quench, say at the initial
time $t=0$ from a frequency $\omega_0$ to the frequency $\omega_1$.

The point in phase space of all
degrees of freedom, denoted previously by $\Gamma$, and which
includes all the bath variables, is supposed not to be accessible.
As {\it available statistical information} we consider the probability
distribution for position $x$ and momentum $p$ at a single arbitrary instant of time
$t$ after the quench.
Then, statistical information on this reduced set of variables at just one
particular time must provide again lower bound for the dissipated
work corresponding to such quench:
\begin{equation} \label{boundxp}
\langle W_{diss} \rangle \geq
 T D(\rho(x,p;t) ||{\tilde{\rho}}(x,-p;t)).
\end{equation}
Below we will elucidate the effect of coarse graining implied in the punctual measurement in time (at time $t$) and, moreover, on a reduction in the number of variables (measuring only $x$, only $p$ or both). Note that we are free to decide what we call the final time of the experiment, hence the choice of the measurement time after the quench is also completely free.

\subsection{Mean dissipated work}

The average work $\langle W_{diss} \rangle$ dissipated at the moment
of the instantaneous quench can be evaluated as follows. The
potential energy of the particle when at a position $x$, is given by
$V_i(x)=m\omega_i^2x^2/2$, where $\omega_i$ is the harmonic frequency, with
$i=0$ and $i=1$ before and after the quench, respectively. The
probability distribution of the position at the moment of the quench is given by
$\rho^{eq}_0(x)=\exp(-V_0(x)/T)/Z_0$ (as before, Boltzmann's constant is absorbed in the temperature for
simplicity of notation). Here $Z_0$, the
normalization constant, is the familiar partition function.
Averaging with respect to this distribution (notation $\langle...\rangle_0$), we
conclude that the average work associated to the quench is given by
$ \langle W\rangle=\langle V_1(x)\rangle_0-\langle
V_0(x)\rangle_0=(T/2)(\omega_1^2/\omega_0^2-1)$. The corresponding
change in free energy is found to be $\Delta
F=-T\ln(Z_1/Z_0)=T\ln(\omega_1/\omega_0)$. Therefore, the total
dissipation in the irreversible instantaneous quench reads
\begin{equation}
\langle W_{diss}\rangle \equiv \langle W\rangle -\Delta F=
\frac{T}{2}\left( \ln \frac{\omega_0^2}{\omega_1^2}+\frac{\omega_1^2}{\omega_0^2}-1
\right).
\end{equation}
Note that the total dissipated work is always positive due to the
irreversible nature of the process.

\subsection{Probability density in forward and backward scenario}

To obtain the bound from the coarse-grained relative entropy
appearing in the r.h.s. of Eq.~(\ref{boundxp}), we need to evaluate
the probability distributions in forward and backward scenario. The
derivation for the backward scenario is very simple. The system
starts at canonical equilibrium with frequency $\omega_1$, and the
quench is performed at the end of the experiment ($t=0$ in forward
time,  which is the final time in the reverse experiment). The
particle is then at canonical equilibrium with respect to the
frequency $\omega_1$ throughout the process, so that
\begin{equation}
\tilde\rho(x,p;t)=\rho_{1}^{eq}(x,p)=
\frac{e^{-(p^2/2m+m\omega_1^2x^2/2)/T}}{Z(\omega_1)}.
\end{equation}
Note that the distribution is even in $p$, namely $\tilde\rho(x,p;t)=\tilde\rho(x,-p;t)$.
Hence the distribution $\tilde{\rho}$ is Gaussian with the following
moments:
\begin{eqnarray}\label{condin1}
& \langle \tilde x\rangle =
\langle \tilde p\rangle =
\langle \tilde{x p}\rangle= 0, & \nonumber \\
& \langle \tilde x^2 \rangle =T/(m\omega_1^2), & \nonumber \\
&\langle \tilde p^2\rangle = mT. &
\end{eqnarray}

One the other hand, in the forward scenario, the initial condition
is canonical with respect to the initial frequency $\omega_0$,
$\rho(x,p;0)=\rho_0^{eq}(x,p)$. At $t=0$ the frequency is
suddenly changed to $\omega_1$ and then kept constant along the
whole process. Therefore, the evolution of the system in the forward
process consists of a relaxation to the new equilibrium state,
$\rho_1^{eq}(x,p)$.
We write the familiar equations of
motion for an underdamped Brownian particle for times $t>0$
\begin{eqnarray} \label{le}
\dot p(t) &=&- m\omega_1^2
x(t)-\lambda \frac{p(t)}{m} +\xi(t), \\
\label{le2}
\dot x(t) &= &\frac{p(t)}{m},
\end{eqnarray}
where $\lambda$ is the friction coefficient, and $\xi$ is Gaussian
white noise with strength determined by the fluctuation dissipation
theorem, $\langle \xi(t)\xi(t')\rangle = 2\lambda T\delta(t-t')$.
The initial condition is stipulated by the fact that prior to the
quench at $t=0$, the system is at equilibrium in a harmonic
potential with strength $\omega_0$, i.e. it is bi-Gaussian with
(compare with Eq.~(\ref{condin1}))
\begin{eqnarray}\label{condin}
& \langle x\rangle_{(t=0)} =
\langle p\rangle_{(t=0)} =
\langle x p\rangle_{(t=0)}= 0, &\nonumber \\
&\langle x^2\rangle_{(t=0)} = T/(m\omega_0^2), &\nonumber \\
& \langle p^2\rangle_{(t=0)} = mT.&
\end{eqnarray}
Since the Langevin equation is linear, the resulting time dependent
probability distribution $\rho(x,p;t)$ remains a Gaussian.
Therefore, it is sufficient to evaluate the ensuing time evolution
of first- and second-order moments. Since there is no shift in the
center  position of the harmonic potential, the average position and
momentum stay equal to zero: $\langle x(t)\rangle =\langle
p(t)\rangle=0$. The second order moments on the other hand obey the
following evolution equations which following directly from the
equations (\ref{le}) and (\ref{le2}):
\begin{eqnarray}\label{eqs}
\frac{d}{dt} \langle x^2\rangle &=& \frac{2}{m}\langle xp\rangle, \nonumber\\
\frac{d}{dt} \langle p^2\rangle &=& -2m\omega_1^2 \langle xp\rangle -\frac{2\lambda}{m}
\langle p^2\rangle +2\lambda T, \nonumber\\
\frac{d}{dt} \langle xp\rangle &=& \frac{1}{m}\langle p^2\rangle -
m\omega_1^2 \langle x^2\rangle -\frac{\lambda}{m} \langle xp\rangle.
\end{eqnarray}
These have to be solved with the above mentioned initial conditions.
One finds:
\begin{equation} \fl \label{eqx2}
\langle x^2\rangle _{t}= \frac{T}{mw_1^2}\left[
1-\frac{\omega}{1-\sigma^2}\; e^{-t\lambda/m}
\left(
\sigma^2/2-(1-\sigma^2/2)\cosh [t\nu]
-\frac{m}{\lambda}\nu\sinh[t\nu]
\right)
\right],
\end{equation}
\begin{equation}\fl \label{eqp2}
\langle p^2\rangle_t = mT \left[   1+
\frac{\sigma^2}{1-\sigma^2} \; \omega \;
e^{-t\lambda/m}
\sinh^2 \left( t \nu/2 \right)
\right],
\end{equation}
\begin{equation} \fl \label{eqxt}
\langle xp\rangle_t = \frac{mT}{\lambda}
\frac{\omega}{1-\sigma^2}
e^{-t\lambda/m}\left[
1-\cosh (t \nu) - \frac{m}{\lambda}\nu\sinh (t\nu)
\right],
\end{equation}
where
\begin{eqnarray}
\omega \equiv \left( \frac{\omega_1}{\omega_0} \right)^2-1,
\; \;
\nu \equiv
\frac{\lambda}{m} \sqrt{1-\sigma^2},
\; \;
\sigma\equiv \frac{2m\omega_1}{\lambda}.
\end{eqnarray}
Note the switch from a monotonously decay ($\nu$ real) to an oscillatory one ($\nu$ imaginary)
of the above solutions for the moments as $\sigma $ crosses the value $1$ from below.

\subsection{Relative entropy}

We are now in position to evaluate the relative entropy (or
Kullback--Leibler distance) between $\rho(x,p;t)$ and
$\tilde\rho(x,-p;t)$.
The relative entropy between the forward and the backward
distribution can be considered as a distance between $\rho(x,p;t)$
and its final equilibrium state $\rho_1^{eq}(x,p)$, only
reached for $t\to \infty$.
Since both densities are Gaussian (and the
backward distribution is even in $p$), the following simple result
is obtained:
\begin{equation} \label{DKL2}
D_{x,p}(t) \equiv D(\rho(x,p;t)||\tilde\rho(x,-p;t))=
 \ln \sqrt{ \frac{{\rm det}\, \tilde C_2}{{\rm det}\,
C_2} } +\frac{{\rm Tr} (\tilde C_2^{-1}C_2)}{2}-1,
\end{equation}
where $C_2$ and $\tilde C_2$ are the covariance matrices of the
forward and backward distributions, respectively
\begin{equation} C_2=\left(
\begin{array}{cc} \langle x^2\rangle_t & \langle xp\rangle_t \\
\langle xp\rangle_t & \langle p^2\rangle_t \end{array}\right),
\end{equation}
\begin{equation} \tilde C_2=\left(
\begin{array}{cc} \langle \tilde x^2\rangle  & \langle \tilde{xp}\rangle\\
\langle \tilde{xp}\rangle & \langle \tilde p^2\rangle
\end{array}\right).
\end{equation}

The above result can be further simplified to
\begin{equation} \label{DKLxp}
D_{x,p}(t)= \frac{1}{2} \left( \ln \frac{\langle \tilde
x^2\rangle \langle \tilde p^2\rangle  }{\langle  x^2\rangle_t
\langle p^2\rangle_t-\langle  xp\rangle_t^2} + \frac{\langle
x^2\rangle_t}{\langle \tilde x^2\rangle}+ \frac{\langle
p^2\rangle_t}{\langle \tilde p^2\rangle} -2   \right).
\end{equation}
From now on,  subindices  in $D$ refer to the variables of
the probability distributions for which the relative entropy is
evaluated.
We also mention the results for the relative entropy of the
probability distribution of only the position
$D(\rho(x;t)||\tilde\rho(x;t))$ and  momentum $
D(\rho(p;t)||\tilde\rho(-p;t))$:
\begin{equation} \label{DKLx}
D_x(t)
=\frac{1}{2}\left( \ln \frac{\langle \tilde x^2\rangle}{\langle  x^2\rangle_t}
 +\frac{\langle  x^2\rangle_t}{\langle \tilde x^2\rangle} -1   \right),
\end{equation}
\begin{equation} \label{DKLp}
D_p(t)=
\frac{1}{2}\left( \ln \frac{\langle \tilde p^2\rangle}{\langle  p^2\rangle_t}
 +\frac{\langle  p^2\rangle_t}{\langle \tilde p^2\rangle} -1   \right).
\end{equation}

\begin{figure} [t]
\begin{center}
\includegraphics[angle=270, width=7.7cm]{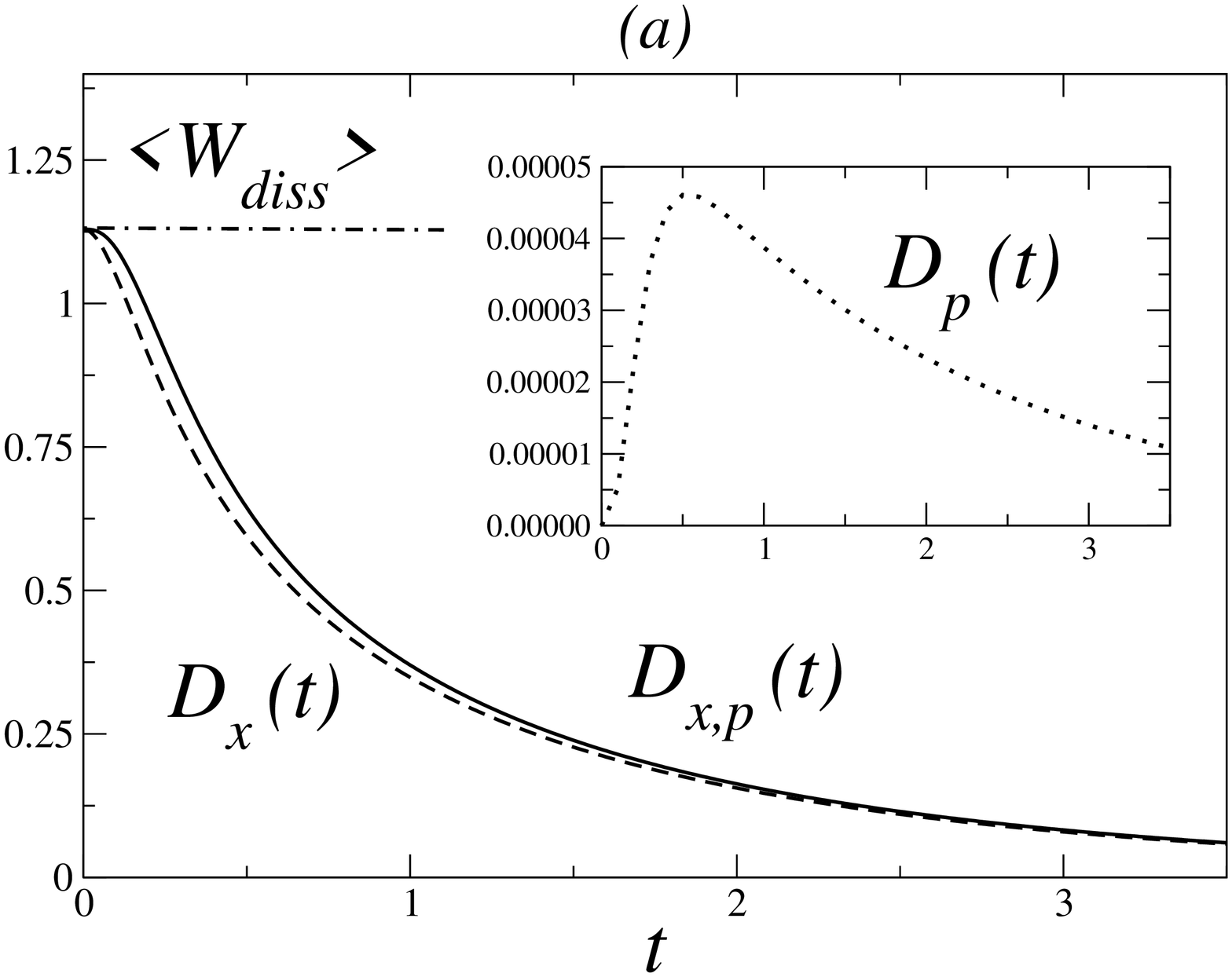}
\includegraphics[angle=270, width=7.7cm]{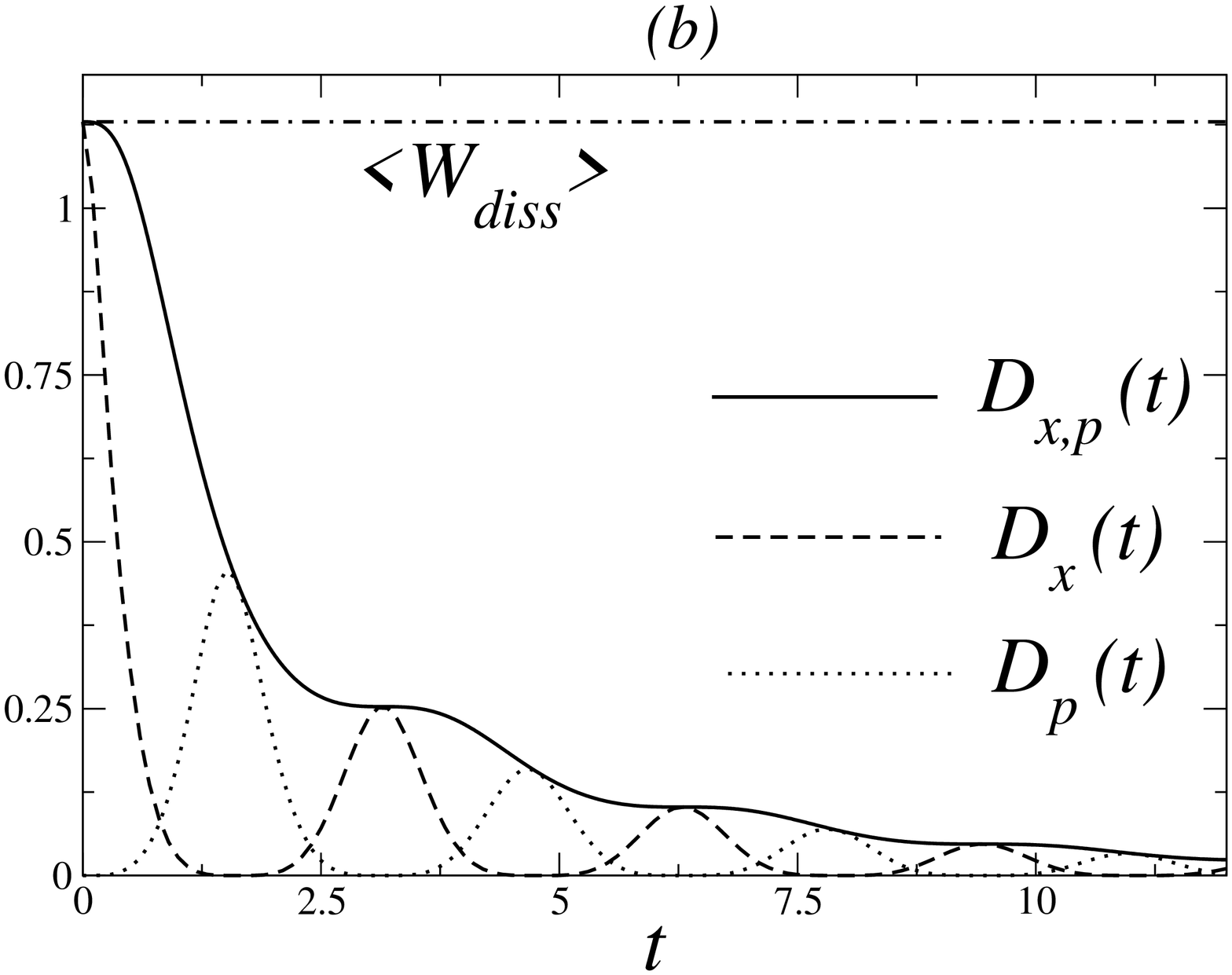}
\caption{(a) Relative entropies $D_{x,p}$, $D_x$ and $D_p$ as a function of time $t$ after the quench in the strongly damped regime ($\sigma<1$): friction dominates inertia. The relative entropies $D_{x,p}$ and $D_{x}$ decay
monotonically and almost coincide, while $D_p$ is very small and
goes through a rise-and-fall.
(b) Relative entropies $D_{x,p}$, $D_x$ and $D_p$ as a function
of time $t$ in the underdamped  regime. Inertia dominates friction
($\sigma>1$) resulting in an oscillatory out of phase decay of the relative entropies $D_x$ and $D_p$. Note that in both cases (a) and (b),  the position
captures the full information on $\langle W_{diss}\rangle$ at $t=0$.  } \label{Dxpfig2}
\end{center}
\end{figure}

With these explicit results (depicted in figure \ref{Dxpfig2}),
we can discuss how well the various
relative entropies capture the dissipation.  We first note that at
the moment of the quench, full information on the dissipation is
captured completely in the statistical information on the position
variable ($D_x(0)=\langle W_{diss}\rangle/T$), while none is
available from the momentum variable ($D_p(0)=0$). This is
consistent with the observation that the position variable is the
only variable which is out of equilibrium at this time.

Furthermore, it is known that the relative entropy between the probability
distribution of a Markov process and its corresponding stationary
state is a strictly decreasing function of time \cite{cover}. 
Hence, $D_{x,p}(t)$ must be a decreasing
function, as we have obtained in our calculation. However, the
relative entropies when only one of the variables is taken into
account exhibit a richer phenomenology.

The behavior is rather different  in the
weakly damped regime than in the strongly damped one. In the strongly damped
case ($\sigma<1$) the relative entropies $D_{x,p}(t)$ and $D_x(t)$ just decay
monotonically with time, see figure ({\ref{Dxpfig2}.a). However, we
obtain a non-monotonous behavior in the relative entropy of the
momentum distribution. This can be explained as follows. The equilibrium
distribution of the momentum does not depend on the frequency of
the oscillator. Therefore, at the quench time, the forward and
backward momentum distributions are identical. However, once the
potential is quenched, the potential energy is not at equilibrium and as a consequence the kinetic energy momentum distribution will depart from
equilibrium to relax back to the same distribution at a later time. As a consequence $D_p(t)$
increases from $D_p(0)=0$, reaches a maximum and decays back to zero
for long time, as can be seen in the inset of figure
(\ref{Dxpfig2}.a). The maximum is however very low, since damping is strong.

We can see a more pronounced and interesting effect in the underdamped case ($\sigma>1$).
The main  results are represented in figure (\ref{Dxpfig2}.b). Note
the oscillatory exchange of information on dissipation between the
position and velocity variables and the decay of the total
information contained in $D_{x,p}(t)$. In particular all the
available information about the dissipation is periodically
contained in one of the single variables, $x$ or $p$, and it dies out (it gets lost in the bath degrees of freedom) as time evolves.

The peculiarities observed in the right figure (\ref{Dxpfig2}.b) can be better
understood by rewriting the relative entropy in (\ref{DKLxp}) as follows:
 \begin{eqnarray} \label{DDD}
D_{x,p}(t)=D_x(t)+D_p(t) +\frac{1}{2}\ln \left( \frac{1}{1-r_t} \right),
\end{eqnarray}
where the correlation coefficient $r_t$ is given by
\begin{equation}
r_t \equiv \frac{\langle xp \rangle_t^2}{\langle x^2\rangle_t \langle p^2 \rangle_t}.
\end{equation}
Since $0\leq r_t \leq 1$, we first note that the last term in the r.h.s of equation (\ref{DDD}) is always positive, hence:
\begin{equation}\label{Dge}
D_{x,p}(t) \geq D_x(t)+D_p(t).
\end{equation}
We conclude that, in the present case, the sum of information on the dissipation gathered
separately from position and momentum is smaller than that from both
variables taken together.
The equality sign in (\ref{Dge}) is realized when $r_{t}=0$, or
$\langle xp \rangle_{t}=0$. Since the variables are Gaussian, the
condition of zero correlation is tantamount to the independency of
position and momentum. From the oscillating analogue of expression  (\ref{eqxt}), one easily verifies that
this occurs at specific times
$t=\frac{2\pi n}{\tilde \nu}$,  where $\tilde \nu= \frac{\lambda}{m}\sqrt{\sigma^2-1}$.

Another feature is that one of the variables, either $x$ or $p$,
 loses all information on dissipation at  another set of specific times.
From equations (\ref{DKLx}) and (\ref{DKLp}) one finds that this
occurs if $\langle x^2 \rangle_t =\langle \tilde x^2 \rangle$ or
$\langle p^2 \rangle_t =\langle \tilde p^2 \rangle$ respectively.
This is in agreement with the more general observation that the
relative entropy of a specific degree of freedom is zero, when, at a
given time, the detailed balance condition holds (i.e., when at that
time the forward and backward distributions are equal).

We conclude that, on the whole, an intricate transfer of information
on dissipation is taken place between position and momentum
of the underdamped Brownian particle. At the same time, the
information on the dissipated work is irreversibly lost by the punctual (one-time) relative entropy of $x$
and $p$ and transfered to the heat
bath variables as time goes by.

\section{Two coupled oscillators: flow of information on dissipation}\label{sec:under2}

To complete the picture, we next consider the case of a harmonically
bound Brownian particle that is coupled, via a second Brownian
particle, to the heat bath. The idea is that by monitoring this second
particle, we are including some information on the heat bath (of
which it is supposed to be part).

The Langevin equations of motion that describe the system read:
\begin{equation}
 m\ddot x_1= -m\omega^2(t) x_1 -K(x_1-x_2),
\end{equation}
\begin{equation}
 m\ddot x_2= -m\omega_0^2 x_2-K(x_2-x_1)+\xi(t) - \lambda \dot x_2 ,
\end{equation}
\begin{equation}
\langle \xi(t)\xi(t')\rangle = 2\lambda  T\delta(t-t').
\end{equation}

Again we consider the quench experiment. For times $t<0$, the
oscillator under consideration, oscillator $1$, is initially at
equilibrium with $\omega_0$. At $t=0$ we perform an
instantaneous quench switching  so that  $\omega(t)=\omega_1$ for
$t>0$. Oscillator $2$ is  kept throughout at the same frequency $w_0$, while, on one hand, linearly coupled  to oscillator $1$ with a strength $K$ and, on the other hand, immersed in the heat bath.
The behavior of  the first moments is trivial:
\begin{equation}
\langle x_1(t)\rangle =\langle p_1(t)\rangle=\langle x_2(t)\rangle =\langle p_2(t)\rangle=0.
\end{equation}
Furthermore, the probability distributions are all Gaussian so we
only need to evaluate the second moments.
Defining $\alpha\equiv K+mw_0^2$ and $\beta \equiv K+mw_1^2$,
they obey the following
set of evolution equations:

\begin{equation}\label{fw4}
\fl \frac{d }{dt} \left( \begin{array}{c}
\langle x_1^2 \rangle \\
\langle p_1^2 \rangle \\
\langle x_1 p_1 \rangle \\
\langle x_2^2 \rangle \\
\langle p_2^2 \rangle \\
\langle x_2 p_2 \rangle \\
\langle  x_1 x_2\rangle \\
\langle x_1 p_2 \rangle \\
\langle x_2 p_1 \rangle \\
\langle p_1 p_2 \rangle
\end{array}
\right) = {\bf A}
\left( \begin{array}{c}
\langle x_1^2 \rangle \\
\langle p_1^2 \rangle \\
\langle x_1 p_1 \rangle \\
\langle x_2^2 \rangle \\
\langle p_2^2 \rangle \\
\langle x_2 p_2 \rangle \\
\langle  x_1 x_2\rangle \\
\langle x_1 p_2 \rangle \\
\langle x_2 p_1 \rangle \\
\langle p_1 p_2 \rangle
\end{array}
\right)
+\left( \begin{array}{c}
0 \\
0 \\
0 \\
0 \\
2 \lambda T \\
0 \\
0 \\
0 \\
0 \\
0
\end{array}
\right),
\end{equation} where the matrix ${\bf A}$ is given by:
\begin{equation}
\fl
 {\bf A}= \left(
\begin{array}{cccccccccc}
0 &  0 & 2/m & 0 & 0 & 0 & 0 & 0 & 0 & 0 \\
0 &  0 & -2\beta & 0 & 0 & 0 & 0 & 0 & 2K & 0 \\
-\beta &  1/m & 0 & 0 & 0 & 0 & K & 0 & 0 & 0 \\
0 &  0 & 0 & 0 & 0 & 2/m & 0 & 0 & 0 & 0 \\
0 &  0 & 0 & 0 & -2\lambda/m & -2\alpha & 0 & 2K & 0 & 0 \\
0 &  0 & 0 & -\alpha & 1/m & -\lambda/m & K & 0 & 0 & 0 \\
0 &  0 & 0 & 0 & 0 & 0 & 0 & 1/m & 1/m & 0 \\
K &  0 & 0 & 0 & 0 & 0 & -\alpha & -\lambda/m & 0 & 1/m \\
0 &  0 & 0 & K & 0 & 0 & -\beta & 0 & 0 & 1/m \\
0 &  0 & K & 0 & 0 & K & 0 & -\beta & -\alpha & -\lambda/m
\end{array}
\right)
\end{equation}

The system can be solved explicitly using the appropriate initial
conditions. However, the analytic expressions are extremely
lengthy. In what follows, we will illustrate the obtained behavior via appropriate figures.

\begin{figure} [t]
\begin{center}
  \includegraphics[angle=270, width=7.7cm]{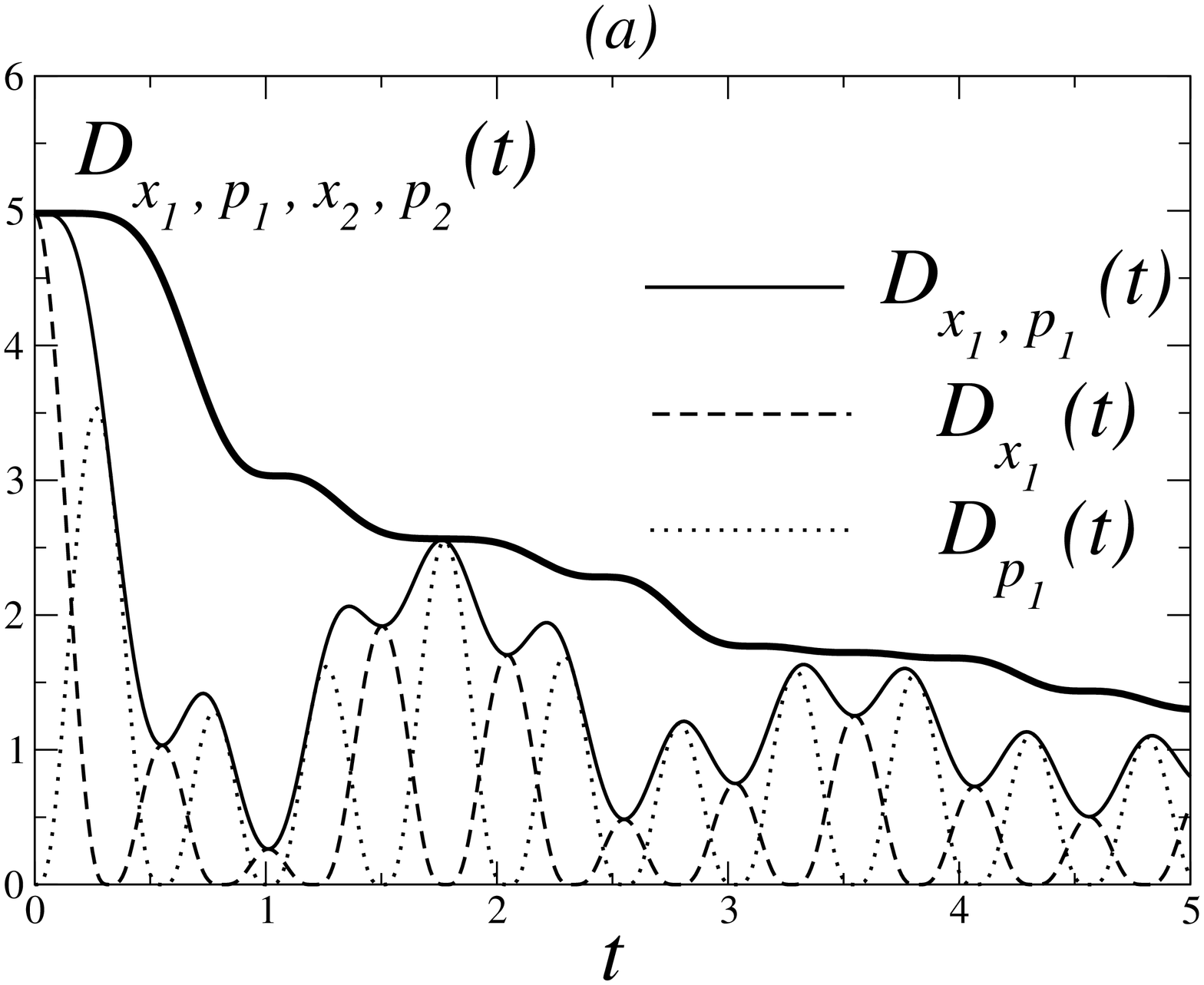}
 \includegraphics[angle=270, width=7.7cm]{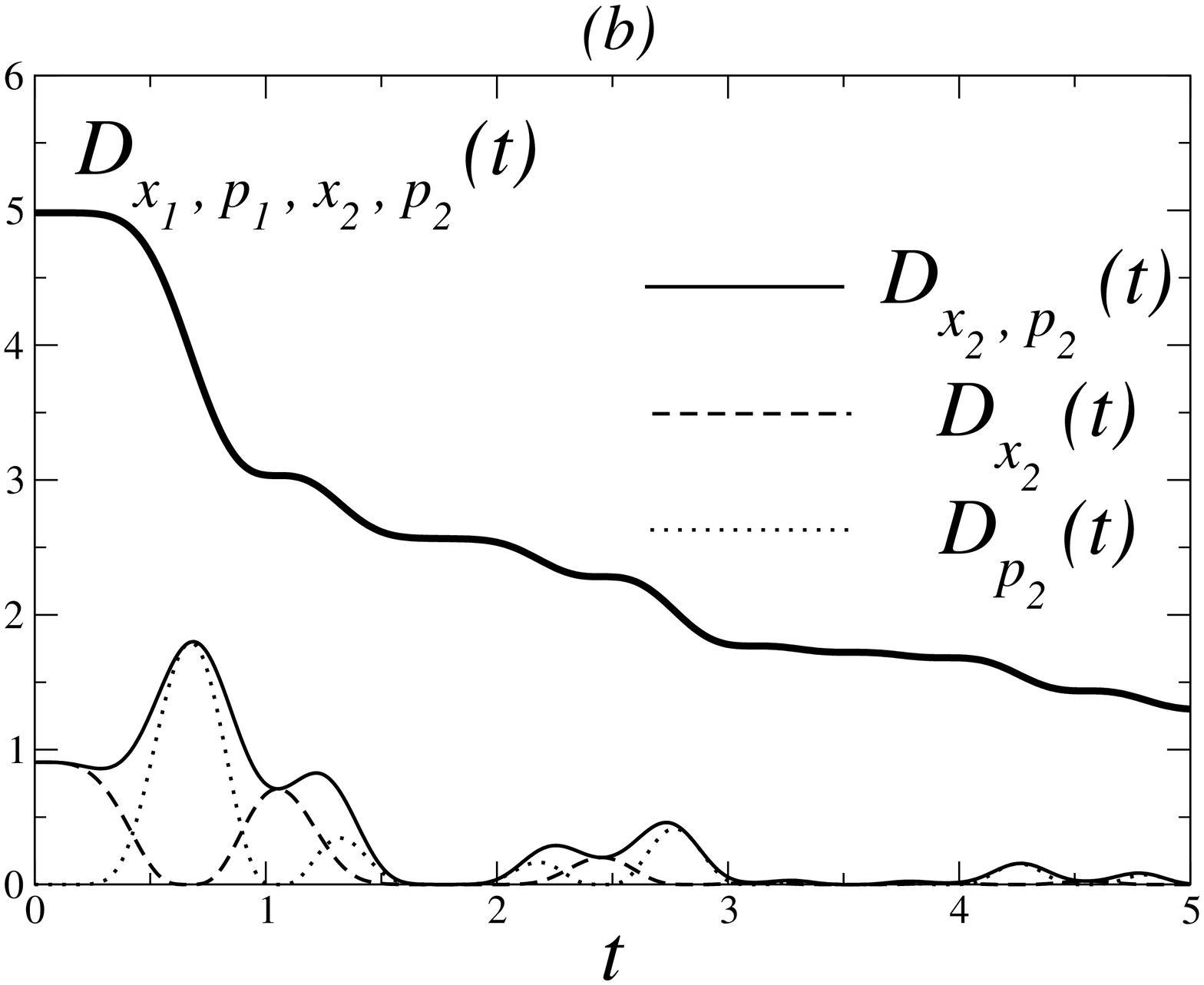}
  \caption{(a) Complex behavior of the relative entropies as a
  function of time for oscillator one. We plot the evolution of
  the total relative entropy of all four variables (positions
  and momenta of both oscillators: $x_1$, $p_1$, $x_2$, $p_2$),
  which decays monotonically and is always an upper bound with respect to any other
  relative entropy accounting for less degrees of freedom. The
 relative entropies pertaining to the first oscillator ($D_{x_1,p_1}(t)$, $D_{x_1}(t)$ and $D_{p_1}(t)$)
   all oscillate in time in an intricate manner.
  (b) Same picture, but for oscillator two. Since the latter
   is directly connected to the heat bath, the relative
  entropy $D_{x_2,p_2}(t)$ is found to decay faster. }
\label{D4x4fig}
\end{center}
\end{figure}

\subsection{Relative entropy}

Since the joint distribution is Gaussian (and the backwards density
even in $p$), the relative entropy involving all four variables
$x_1,p_1,x_2$ and $p_2$ can be compactly expressed in terms of the
covariance matrices $C_4$ and $\tilde C_4$:
\begin{eqnarray} \label{DKL4}
D_{x_1,p_1,x_2,p_2}(t)=
\frac{1}{2} \left[ \ln\left( \frac{{\rm det}\, \tilde C_4}{{\rm det}\,
C_4}\right) +{\rm Tr} (\tilde C_4^{-1}C_4)-4\right].
\end{eqnarray}
The latter  are the following four-by-four symmetric matrices
\begin{equation} C_4=\left(
\begin{array}{cccc}
\langle x_1^2\rangle_t & \langle x_1p_1\rangle_t & \langle x_1x_2\rangle_t & \langle  x_1 p_2\rangle_t\\
\langle x_1 p_1\rangle_t & \langle p_1^2\rangle_t & \langle p_1 x_2\rangle_t & \langle p_1 p_2\rangle_t\\
\langle x_1 x_2\rangle_t & \langle p_1 x_2\rangle_t & \langle x_2^2\rangle_t & \langle x_2 p_2\rangle_t\\
\langle x_1 p_2\rangle_t & \langle p_1 p_2\rangle_t & \langle x_2 p_2\rangle_t & \langle p_2^2\rangle_t
\end{array}\right).
\end{equation}
Regarding the covariance matrix corresponding to the backwards excursion, we explicitly find
\begin{equation} \tilde C_4=\left(
\begin{array}{cccc}
\frac{\alpha T}{Kmw_1^2+\beta mw_0^2}
 & 0 & \frac{ K T}{Kmw_1^2+\beta mw_0^2}
  & 0 \\
0 & mT & 0 & 0 \\
\frac{ K T}{Kmw_1^2+\beta mw_0^2} & 0 & \frac{ \beta T}{Kmw_1^2+\beta mw_0^2} & 0 \\
0 & 0 & 0 & mT
\end{array}\right).
\end{equation}

From the above results, we can derive the relative entropy of {\it all available} degrees of freedom of the system (both
positions $x_1$ and $x_2$, and momenta $p_1$ and $p_2$). While these are the pertinent variables to evaluate the dissipated work, when measured along the whole time track, a single time measurement as performed here again
represents a coarse-graining. Only when performed at the moment of the quench does it contain full information. When considering times $t >0$, information on the dissipation will
 flow and get irreversibly lost to the bath variables. This is similar to the situation discussed in the overdamped case

Similarly to the underdamped oscillator case, one can also explore
the behavior of the relative entropies of all possible combinations
of all 4 degrees of freedom  $x_1$, $p_1$, $x_2$ and $p_2$. Several
of such combinations are plotted in figure \ref{D4x4fig}. First note
that, as explained before, the relative entropy of the whole system,
$D_{x_1,p_1,x_2,p_2}(t)$, decays monotonically in time. Then, the
relative entropy of oscillator $1$, $D_{x_1,p_1}(t)$,  is
oscillating below the former. Both entropies for position and
momentum alone, transfer information periodically and are modulated
by $D_{x_1,p_1}(t)$. Note that the position of the first oscillator
captures the whole  dissipation at the moment of the quench,
\begin{equation}
D_{x_1}(0)=D_{x_1,p_1}(0)=  D_{x_1,p_1,x_2,p_2}(0).
\end{equation}

The novelty in this case is that oscillator $1$ is not directly in
contact with  the heat bath, but rather to oscillator $2$, whose
relative entropies we now comment on. First, we see that the
relative entropies are significantly smaller than those in
oscillator $1$. Oscillator 2 receives the information on the
dissipated work from the quench only indirectly through its coupling
to $1$. Furthermore, while it ``bounces'' back some of this
information to $1$ it also irreversibly loses information to the
bath variables.

\begin{figure} [t]
\begin{center}
 \includegraphics[angle=270, width=7.7cm]{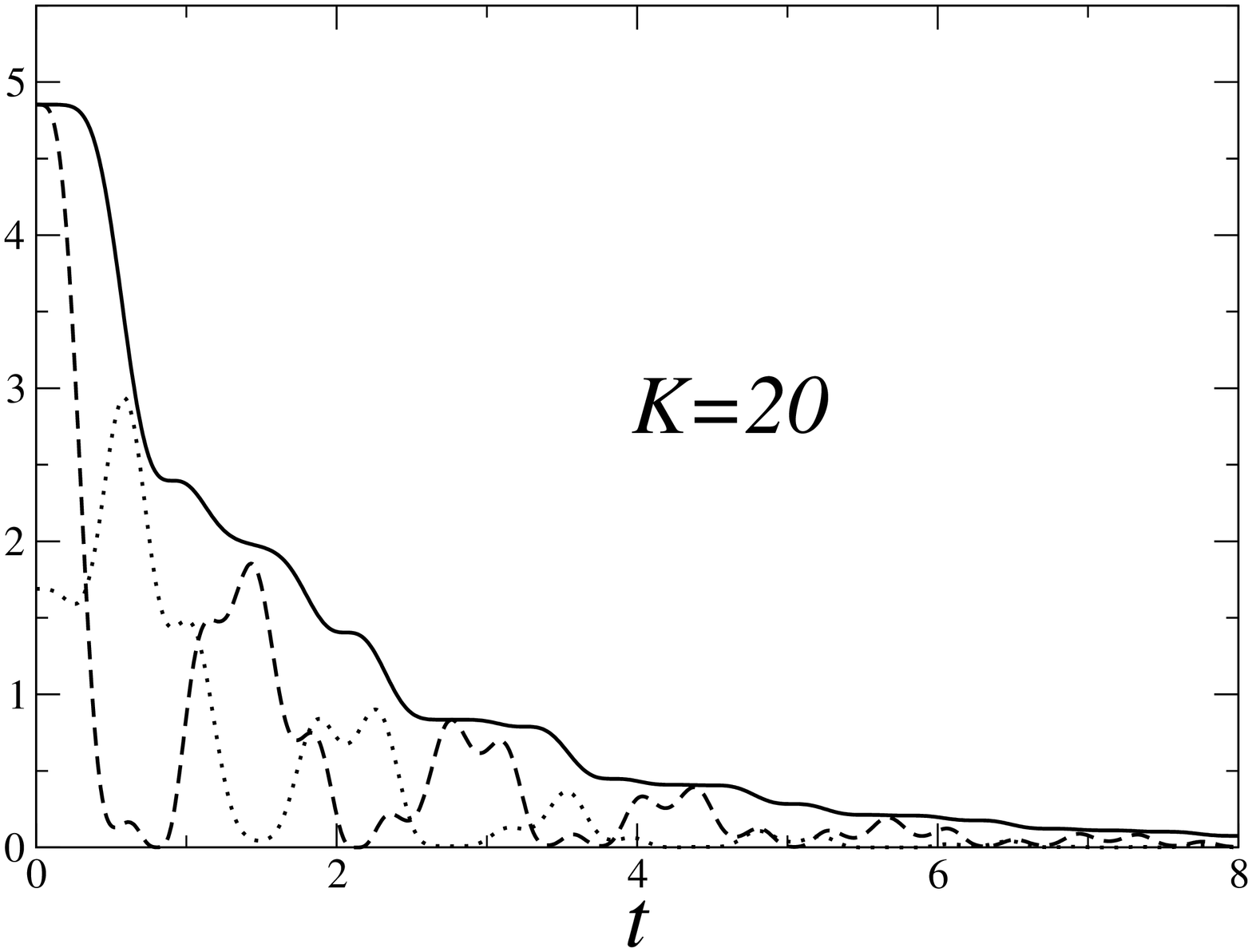}
  \includegraphics[angle=270, width=7.7cm]{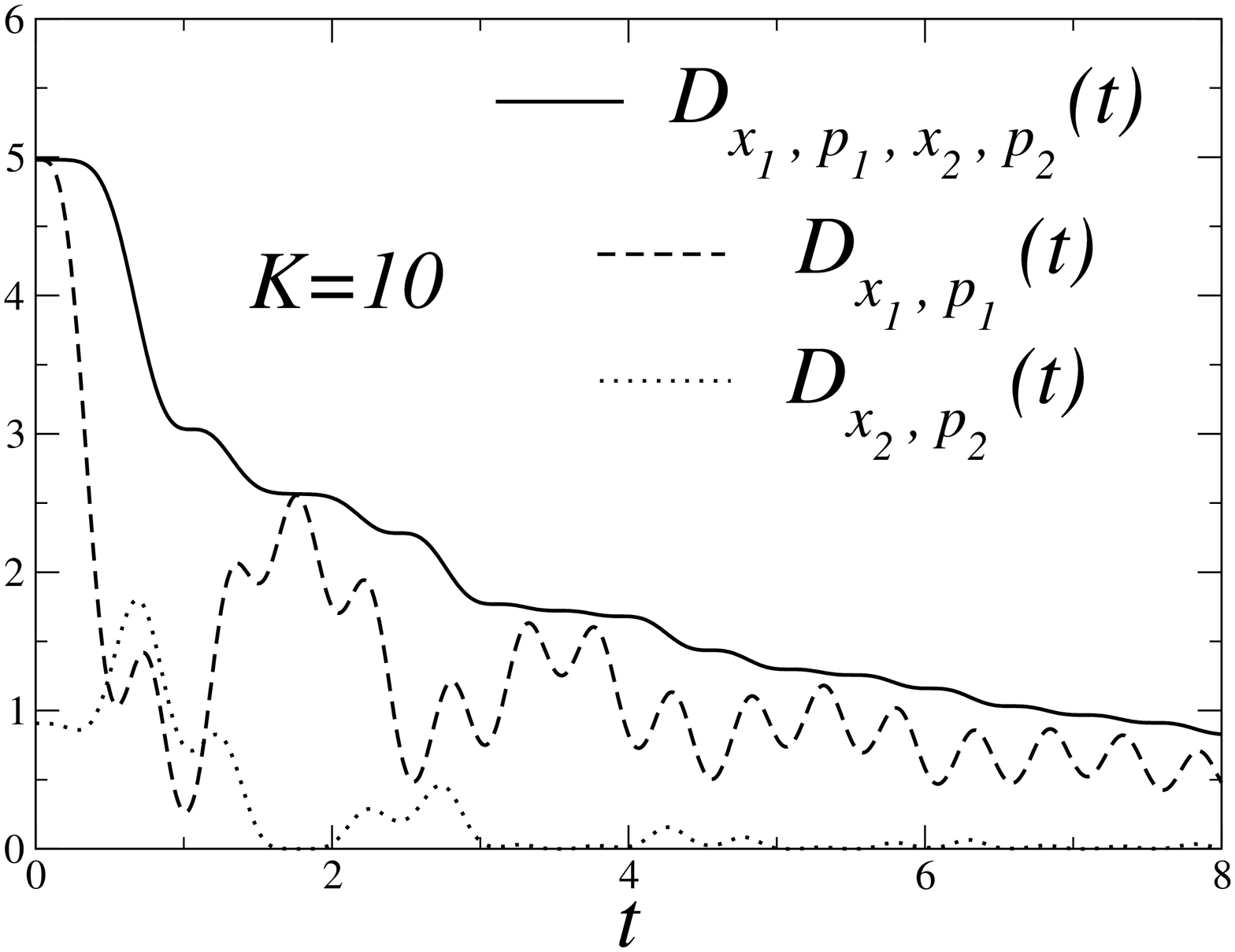}
 \includegraphics[angle=270, width=7.7cm]{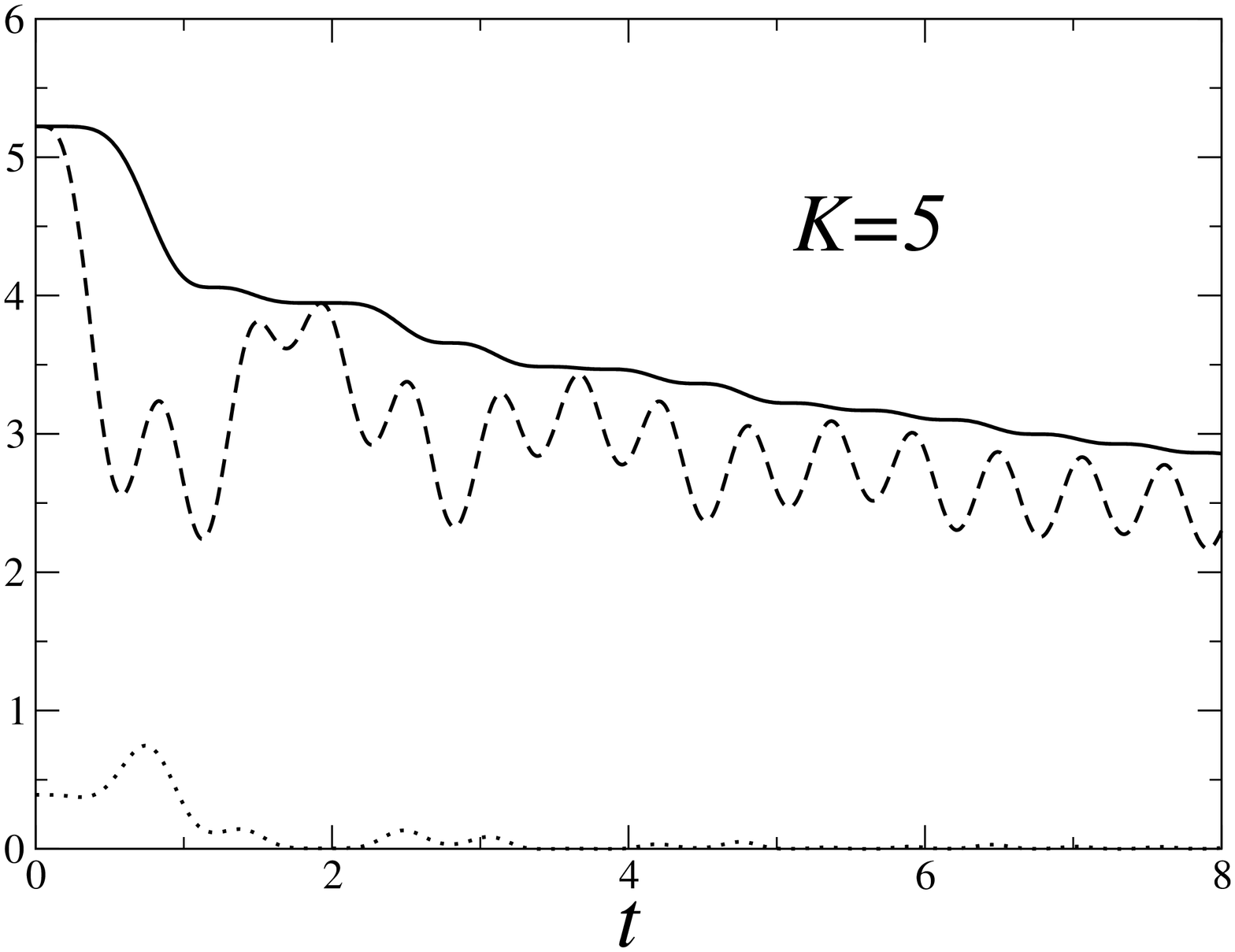}
  \includegraphics[angle=270, width=7.7cm]{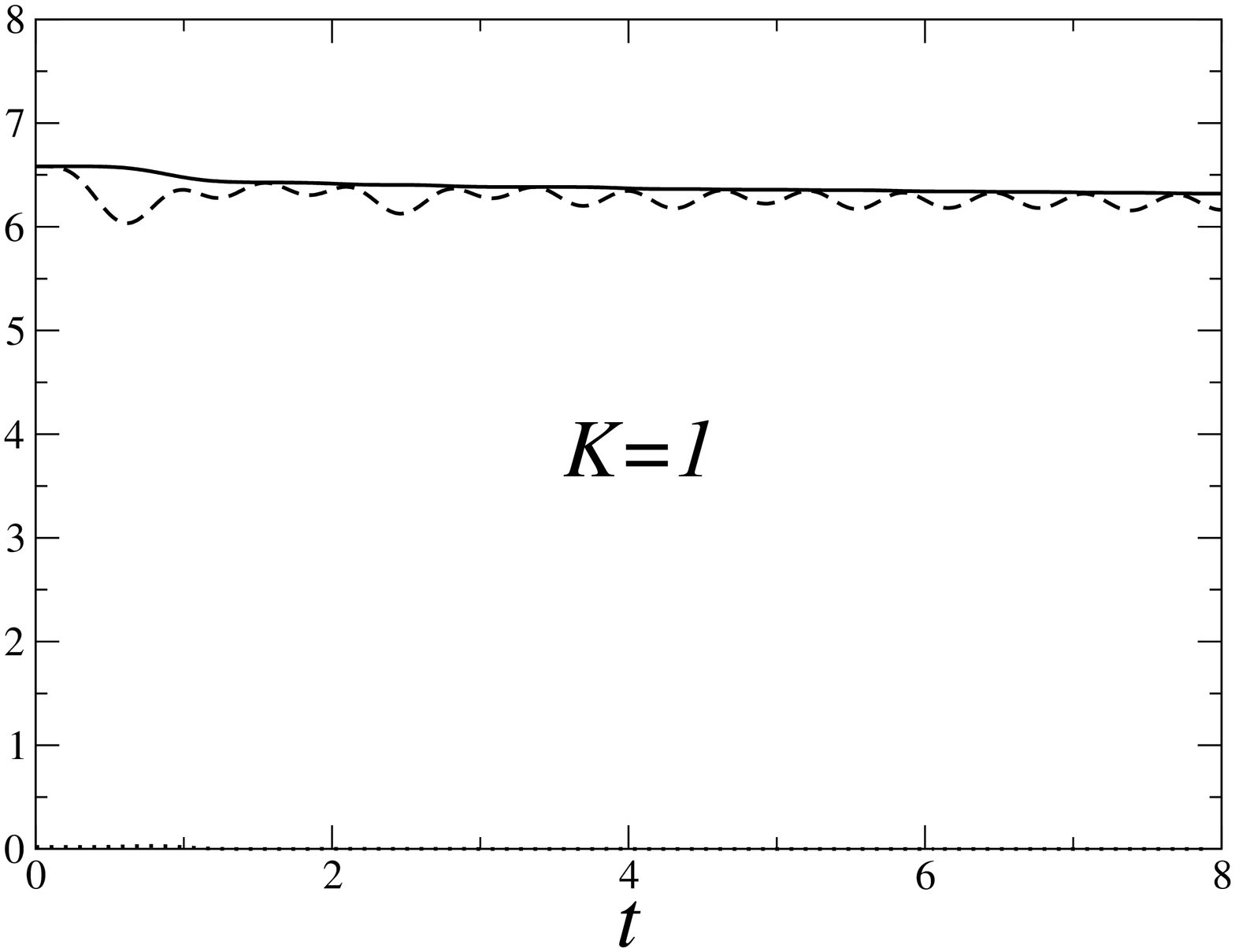}
 \caption{Time evolution of the relative entropies for the whole system,
 $D_{x_1,p_1,x_2,p_2}(t)$, oscillator one,
  $D_{x_1,p_1}(t)$, and oscillator two, $D_{x_2,p_2}(t)$.
  Each plot shows a different behavior as the coupling
  constant $K$ between the oscillators is changed. This illustrate
  a subtle mechanism of information transfer due to  correlations
  of the degrees of freedom. }
\label{Dkexplore}
\end{center}
\end{figure}

Regarding the plateau for the relative entropy appearing at short times in figure (\ref{D4x4fig}), one observes that most of the effect of the dissipative process still resides
inside the system formed by the two particles. In fact, oscillator $1$ keeps
much of this information  while
slowly transferring it to oscillator $2$. Some of it can come back but a part of it
is lost to the heat bath.
This behavior is illustrated in figure (\ref{Dkexplore}) by varying the coupling constant $K$ that
connects both oscillators.  Note that for $K=5$, the
relative entropy of oscillator $2$ is almost zero but yet there is a
considerable difference between $D_{x_1,p_1,x_2,p_2}(t)$ and
$D_{x_1,p_1}(t)$. Therefore, while oscillator $2$ is "close to equilibrium",
its correlation with oscillator $1$ still carries relevant
information on the irreversible quench.

\section{Conclusions}

We have derived a microscopically exact expression for the dissipative work along an arbitrary process starting in equilibrium for systems described by stochastic dynamics. As was anticipated in earlier work in the literature, we find that dissipation is proportional to the relative entropy between the probability distributions of forward and backward trajectories, respectively. In other words, dissipation is related to our ability to distinguish the arrow of time. Furthermore, the expression in terms of relative entropy gives rise to lower bounds if only partial information on the trajectories is available.

In combination with Crook's theorem \cite{crooks,work5}, we have shown that a single functional of the path, namely the work itself, provides an exact and full assessment of dissipation. In other words, work "exhausts" the arrow of time: if we know the statistical properties of the work done along the forward and the backward process, no further information can help us to better distinguish between the two.

We have discussed various scenarios to illustrate how dissipation can be bounded from below on the basis of reduced information.  First, when the information about the continuous trajectory of the
system is reduced to a finite number of measurements, our analysis has shown that the resulting relative entropy provides reasonably accurate bounds for the dissipation, even with only a small number of intermediate measurement points. This result could be specially useful in real experiments where trajectories are recorded at finite sampling rates. Second, we have analyzed the effect of considering a subset of variables instead of a detailed description of the system in a quench process. In this case, the time-arrow information, concentrated in the single position variable immediately after the quench, is subsequently transferred  to the thermal bath and the other variables.
Of special interest is the case of two oscillators, the first one undergoing a quench of its frequency and the second one in contact with a thermal bath.
One would expect that the information contained in the first oscillator would be transferred to the second one before getting lost in the thermal bath. However, our analysis calls into question this naive picture, as we have shown that the oscillator coupled to the thermal bath is the first to thermalize. This result indicates that ``reversibility"тт is transferred from the thermal bath to the quench point, instead of ``irreversibility" being transferred in the opposite direction. The generalization of our analysis to long chains of oscillators will help to further elucidate this issue.

\ack We acknowledge financial help from the Ministerio de Educacion
y Ciencia (Spain) under Grants FPU-AP-2004-0770 (A. G--M.) and
MOSAICO (JMRP), and from the FWO Vlaanderen.

\section*{Appendix}

We studied the effect of coarse-graining in time for the harmonic
potential. However, the same procedure can be applied for an
arbitrary potential, at least in the limit of a infinitely large
number $n$ of time measurements. Indeed the Gaussian ansatz
remains valid for any potential for {\it for short time increments},
since the propagator of the Fokker-Planck equation is then always
Gaussian \cite{risken}. Therefore, equations (\ref{pF}) and
(\ref{pB}) are completely general in this case. The only difference is
the expression for the first and second moments, but even this we know, since
\begin{equation}
x_{i+1}(t_i+\Delta t) = x_{i}(t_i) -V'(x_i,t_i)\Delta t + \xi(t_i) \Delta t^{1/2}+\mathcal{O}\left((\Delta t)^2\right).
\end{equation}
On the whole, for small $\Delta t$, the Fokker-Planck equation of
such transition probabilities can be solved giving rise to
\cite{gardiner}
\begin{equation}
p(x_{i+1}, t+\Delta t| x_{i},t)= \frac{1}{\sqrt{4\pi T \Delta t}}
 \exp\left[ -\frac{
(x_{i+1}-  x_{i} + V'(x_i,t) \Delta t )^2
  }{4 T \Delta t}\right].
\end{equation}

Now we recall equation (\ref{sumIn}), and we find that the first term in the r.h.s. gives
\begin{equation} \label{firstn}
T\left\langle  \ln \frac{p^{eq}_0(x_0)}{p^{eq}_1(x_f)} \right\rangle =
\langle \Delta V \rangle - \Delta F,
\end{equation}
where $\Delta V=V(x_f,t_f)-V(x_0,t_0)$. After substitution of the
conditional probabilities and some algebra, the second term in the r.h.s. of
(\ref{sumIn}) yields to
\begin{equation} \label{secondn}
T\sum^{n-1}_{i=0} \left\langle \ln \frac{p(x_{i+1}|x_i)}{\tilde{p}(x_i|x_{i+1})}
\right\rangle=\big\langle \sum^{n-1}_{i=0}(A+B)\big\rangle,
\end{equation}
where
\begin{equation}
A=-\Delta t \frac{(x_{i+1}-x_{i})}{\Delta t}\frac{[V'(x_{i+1},t_{i+1})+V'(x_i,t_i)]}{2},
\end{equation}
\begin{equation}
B=\Delta t \frac{  [V'(x_{i+1},t_{i+1})]^2-[V'(x_i,t_i)]^2 }{4}.
\end{equation}
Remember that this result is valid for small $\Delta t$, or
conversely, for big $n$. In the limit of $n\to \infty$, we find  to
lowest order in $ \Delta t$ that $A=-\Delta t \;\dot{x} V'(x,t)$ and $B=0$,
hence:
\begin{equation} \label{limitQ}
\lim_{n\to\infty} \sum^{n-1}_{i=0}(A+B)=-\int_{t_0}^{t_f}dt \dot{x}V'(x,t).
\end{equation}
The last integral is the heat $Q$ associated to a specific stochastic
trajectory, cf. Eq. (\ref{1stLaw}). Then, relations (\ref{sumIn}),
(\ref{firstn}), (\ref{secondn}) and (\ref{limitQ}), together with
the conservation of energy, imply that
\begin{equation}
I_{n\to \infty} = \langle \Delta V \rangle - \Delta F+ \langle Q \rangle=
 \langle  W \rangle - \Delta F= \langle  W_{diss} \rangle.
\end{equation}
This is a pedestrian path integral method to show that, for a
general time dependent potential,  the relative entropy of the
distributions of the forward and backward paths (considering also their initial ensemble probabilities) is equal to the
dissipated work.

\end{document}